\documentclass{flbook12}
\usepackage{amsmath,amssymb,amsfonts}
\usepackage{epsfig} %Replace with {graphics}?
\usepackage{natbib}

\usepackage{latexsym}
\usepackage{makeidx}
\makeindex

\title{From\\ {\bf The Global Circulation of the Atmosphere}\\Schneider \& Sobel eds. \\Princeton University Press 2007}

\begin{document}
\maketitle
\mainmatter
\chapter{On the Relative Humidity of the Atmosphere}

{\hspace*{18pt}\sffamily{\large Raymond T. Pierrehumbert\\\hspace*{18pt}Helene Brogniez\\\hspace*{18pt}Remy Roca}}

\section{Introduction}

Water is a nearly miraculous molecule which enters into the operation of
the climate system in a remarkable variety of ways.  A wide-ranging
overview can be found in \citet{InsightHydrology}.  In the present
chapter, we will be concerned only with the radiative effects of
atmospheric water vapor on climate, and with the kinematics of how the
water vapor distribution is maintained.  These effects are important
because feedbacks due to changes in atmospheric water vapor amplify the
climate system's response to virtually all climate forcings, including
anthropogenic and natural changes in $CO_2$, changes in solar
luminosity, and changes in orbital parameters. In contrast to cloud
feedbacks, which differ greatly amongst general circulation models,
clear sky water vapor feedback is quite consistent from one model to
another. Essentially all general circulation models yield water vapor
feedback consistent with that which would result from holding {\it
relative humidity} approximately fixed as climate changes
\citep{HS2000,ColmanMcAvaney97}.  This is an emergent property of the
simulated climate system; fixed relative humidity is not in any way
built into the model physics, and the models offer ample means by which
relative humidity could change.

Why should models with such diverse representations of moist processes
yield such similar results when it comes to water vapor feedback? The
answer to this question has a considerable bearing on the extent to
which one can trust models to faithfully reproduce the water vapor
feedback occurring in nature, and in particular in climates which
haven't yet been directly observed.  There is much indirect evidence
that the water vapor feedback in models is correct, and indeed no
compelling reason has emerged to doubt it.  Nonetheless, it has proved
difficult to articulate cleanly and convincingly from basic principles
exactly why one should have confidence in this aspect of the models. If
the atmosphere were saturated at all levels, understanding water vapor
feedback would offer few challenges. The difficulty arises from the
prevalence of highly unsaturated air in the atmosphere. To make progess,
one needs better conceptual models of the factors governing the
distribution of subsaturated air in the atmosphere. Once mechanisms are
understood, prospects for determining whether key processes operate in
the same way in models as in the real world become brighter.

Another reason for seeking a better mechanistic understanding of the
relative humidity distribution is that such an understanding is a
prerequisite for credible incorporation of water vapor feedback in
idealized climate models. Despite advances in computer power, energy
balance models and their somewhat embellished cousins the
"intermediate complexity models" continue to have an important role in
exploration of new ideas. When simulations of tens or hundreds of
thousands of years are called for, they are indispensible.
In thinking about representation of the hydrological cycle in
idealized models, one must distinguish between the requirements of
modelling water for such purposes as computing precipitation
and latent heat transport, and the requirements of modelling water
for the purpose of determining its radiative impacts. The former
is a challenging enough task, but involves primarily low-altitude
processes where most of the atmosphere's water is found. The latter
is yet more formidable, as it requires modelling the relatively small
proportion of water found in the upper reaches of the troposphere. 
Even in idealized models which attempt some representation of
the hydrological cycle \citep{Climber,UVicModel}, the dynamic
influences on water vapor feedback are generally neglected.

The aim of this chapter is to present some simple but quantifiable ideas
concerning the way in which the atmosphere's population of unsaturated air
emerges from the interplay of transport and condensation.  The key insight
is to think of the water content of a parcel of air as resulting from a
stochastic process operating along air parcel trajectories. 
The various incarnations of this idea which will be discussed in this chapter
account for the prevalence of dry air, offer some insight as to how the
relative humidity distribution may change in response to a changing
climate, and yield some diagnostic techniques that can be used as a basis
for comparing moisture-determining processes between models and the real world.
Through detailed analysis of an idealized version of the stochastic problem, we
expose some of the mathematical challenges involved in formulating moisture
parameterizations for idealized climate models, and suggest a possible
approach to stochastic modelling of moisture. The class of theories
we consider here deals only with those aspects of moisture that can be
inferred on the basis of large scale wind and temperature fields such as
can be explicitly resolved by general circulation models and global
analysis/assimilation datasets.  This is only a piece of the puzzle, but
we think it is a large piece.  The whole complex of important
issues concerning the way deep convection injects moisture into the
atmosphere, and the sensitivity of such processes to microphysics
and parameterization (e.g. \citet{TompkinsEmanuel2000}; see also Chapters 3 and 7)
is left untouched for the moment, but must at some point be brought back into the picture.

We begin by reviewing some basic material concerning the radiative
importance of water vapor in Sections \ref{sec:BasicProperties} and
\ref{sec:Importance}. The essential ideas governing the generation of
subsaturated air are introduced in Section \ref{sec:HowSaturated}. A
simple class of models embodying these ideas is analyzed in Section
\ref{sec:SimpleModels}, where we also analyze a more conventional
diffusive model for the sake of comparison and contrast. In Section
\ref{sec:RealTrajec} the concepts are applied to diagnosis of moisture
dynamics in the present climate, and in GCM simulations of warming in
response to elevation of $CO_2$.

\section{Basic properties of water vapor}
\label{sec:BasicProperties}

There are two basic aspects of the physics of water vapor that give
it a distinguished role in determining the sensitivity of the Earth's
climate.  The first aspect is common to all substances that undergo
a phase change: the maximum partial pressure of water vapor that can
be present in a volume of atmosphere in thermodynamic equilibrium
is a strongly increasing function of temperature.  This maximum
is known as the {\it saturation vapor pressure}, $e_s$, and is governed
by the Clausius-Clapeyron relation.  For water vapor at modern 
terrestrial temperatures, the Clausius-Clapeyron relation is
well approximated by the exponential form
\begin{equation}
e_s(T) = e_s(T_o) \exp^{-\frac{L}{R_v} (\frac{1}{T} - \frac{1}{T_o})}
\label{eqn:ClausiusClap}
\end{equation}
where $L$ is the latent heat of the appropriate phase transition 
(vapor to liquid at warm temperatures, vapor to solid at sufficiently
cold temperatures), $R_v$ is the gas constant for water vapor, and
$T_o$ is a reference temperature.
At the freezing point, $e_s$ is $614 Pa$ or $6.14mb$; $L/R_v = 5419K$ for
condensation into liquid, and $6148K$ for condensation into ice.
The formula predicts a very strong dependence of saturation vapor pressure
on temperature.  At 300K, $e_s$ rises to $3664 Pa$, whereas at 250K $e_s$ drops to a mere $77 Pa$. 
Once the partial pressure of water vapor reaches the saturation vapor
pressure, any further addition of water vapor will lead to condensation
sufficient to bring the vapor pressure back down to saturation. The condensed
water may remain in suspension as cloud droplets, or it may aggregate and 
be removed from the atmosphere in the form of precipitation.  Because
specific humidity is a materially conserved quantity until
condensation occurs, it is convenient to define the {\it saturation specific humidity},
$q_s$, which is the ratio of the mass of water vapor that could be held in
saturation to the total mass of saturated air.  In terms of
saturation vapor pressure, $q_s = .622 e_s(T)/(p_a + .622 e_s(T))$, where $p_a$ is the
partial pressure of dry air.  Condensation occurs whenever the specific
humidity $q$ exceeds $q_s$.
\footnote{$q/q_s$ is not precisely equal to the relative humidity, which is
defined as the ratio of the mixing ratio of water vapor to the saturation
mixing ratio (equivalently, the ratio of the partial pressure of water vapor
to the saturation partial pressure). However, for the purposes of this
chapter, where water vapor is assumed to be a minor constituent of the atmosphere,
mixing ratio and specific humidity can be regarded as practically interchangeable.}

The Clausius-Clapeyron relation provides a powerful constraint on the behavior
of water vapor, but it is not at all straightforward to tease out the
implications of this constraint for climate, for the reason that it
only gives an {\it upper bound} on the water vapor content for any given
temperature, and tells us nothing about how closely that bound might be
approached. 

The second key aspect of water vapor is that it is a potent greenhouse
gas. Like most greenhouse gases, the effect of water vapor on Outgoing Longwave Radiation
(OLR) is approximately logarithmic in specific humidity, once the concentration
is sufficiently large to saturate the principal absorption bands.  The logarithmic
effect of water vapor is somewhat more difficult to cleanly quantify than is
the case for well-mixed greenhouse gases like $CO_2$, but if one
adopts a base-case vertical distribution and changes water vapor by multiplying
this specific humidity profile by an altitude-independent factor, one finds
that each doubling of water vapor reduces OLR by about $6 W/m^2$ \citep{ChapmanWater}.
This is about 50\% greater than the sensitivity of OLR to $CO_2$.  A comprehensive
analysis of the latitude and altitude dependance of the sensitivity of OLR to
water vapor can be found in \citet{HS2000}.  The idea that small quantities
of water vapor can have a lot of leverage in climate change has a fairly
long history, and is now widely recognized.  Water vapor feedback was
included in the very first quantitative calculations of $CO_2$-induced 
warming by Arrhenius, and the importance of water vapor aloft was implicit
in such calculations, as it was in the first comprehensive radiative-convective
simulation of the problem by \citet{MW67}. However, consciousness
of the importance of free tropospheric humidity, and of the difficulty of
understanding it, did not really awaken until the early 1990's.  Significant
early work on the subject appeared in \citet{SodenBretherton} and \citet{Shine91},
partly in response to some salutary,if ultimately easily dismissed, skepticism
about global warming expressed in \citet{Coolness}.
Related points have been discussed in \citet{spenbras97,ChapmanWater,HS2000},
among others. 

One important consequence of the logarithmic dependance is that the relatively small
amount of water vapor aloft nonetheless has a great influence on the radiation 
budget. Another consequence is that the degree of dryness of dry air also has
a great effect on the radiation budget:  Reducing free tropospheric humidity
from 5\% to 2.5\% in a dry zone has nearly the same radiative impact as
reducing humidity from 40\% to 20\%, even though the former reduction involves
a much smaller quantity of water. 
A related consequence of the logarithmic dependence is that the radiative effect
of water vapor cannot be accurately determined on the basis of the mean specific humidity
alone.  Fluctuations in specific humidity, e.g. fluctuations at scales smaller
than resolved in a climate model, affect the OLR to some extent.  If we denote
the specific humidity by $q$ and time or space averages by angle brackets, then 
$\langle \log q \rangle \ne \log \langle q \rangle$. Specifically, writing 
$q = \langle q \rangle + q'$,
\begin{equation}
\langle\log(\langle q \rangle (1 + \frac{q'}{\langle q \rangle}))\rangle
  \approx \log(\langle q \rangle) - \frac{1}{2} \langle (\frac{q'}{\langle q\rangle})^2 \rangle
    < \log \langle q \rangle
\end{equation}
Since OLR is proportional to $-\log q$, fluctuations in water vapor {\it increase}
the OLR and have a cooling effect. Having some very dry air and some very moist air
allows more infrared cooling than would the same amount of water spread uniformly
over the atmosphere.   To provide some idea of the magnitude of this effect, consider
a region of the atmosphere within which the temperature is horizontally
homogeneous, and where the relative humidity is 50\% everywhere.  Next, increase the 
humidity in one half of the region to 87.5\%, while keeping the total water
in the system constant. This requires a reduction of the rest of the region to
12.5\% relative humidity. One not-quite doubles the humidity in half of the region,
while reducing the humidity in the rest by a factor of 1/4, yielding a net
increase of OLR of $3.6 W/m^2$, based on the sensitivity factor given above.
If we make the dry air still drier, reducing it to 6.25\% while increasing
the moist air to 93.75\%, the OLR increase relative to the uniform state
becomes $11 W/m^2$.
%**CORRIGENDUM:  Arithmetic in above paragraph is not quite right. See corrigendum.

Water vapor affects climate sensitivity through its effect on the slope
of $OLR$ vs surface temperature.  Following \citet{HS2000}, the change $\Delta T$
in temperature caused by a change $\Delta F$ in radiative forcing can
be written $\Delta T = \Lambda \cdot \Delta F $, where the sensitivity coefficient is
given by
\begin{equation}
\Lambda = \frac{1}{(1-\beta)\frac{\partial}{\partial T} OLR |_{\text{specific humidity fixed}}}
\end{equation}
If free tropospheric water vapor increases with temperature, the additional
greenhouse effect reduces $OLR$ compared to what it would be with water
vapor fixed, whence the water vapor feedback factor $\beta$ is positive,
and climate sensitivity is increased.

General circulation model simulations of the change in the present climate
caused by increasing $CO_2$ universally show {\it polar amplification},
in the sense that the temperature increase at high latitudes is greater than
that at low latitudes.  Observations of climate change over the past
century seem to conform to this expectation.  It is important to note
that water vapor feedback does not contribute to polar amplification.
In fact, for the present pole to equator temperature range, water vapor
feedback makes the sensitivity $\Lambda$ {\it smaller} at cold temperatures
than at warm temperatures (see \citet{InsightHydrology}), and hence would lead to 
tropical rather than polar amplification.  Certainly, ice albedo feedback
plays a role in polar amplification, but dynamical heat transport and
clouds may also contribute.  These feedbacks must be sufficiently strong
to overcome the tendency of water vapor feedback to put the greatest warming
in the tropics.

\section{The importance of water vapor to the radiation budget}
\label{sec:Importance}
\begin{figure}
\epsfig{file = 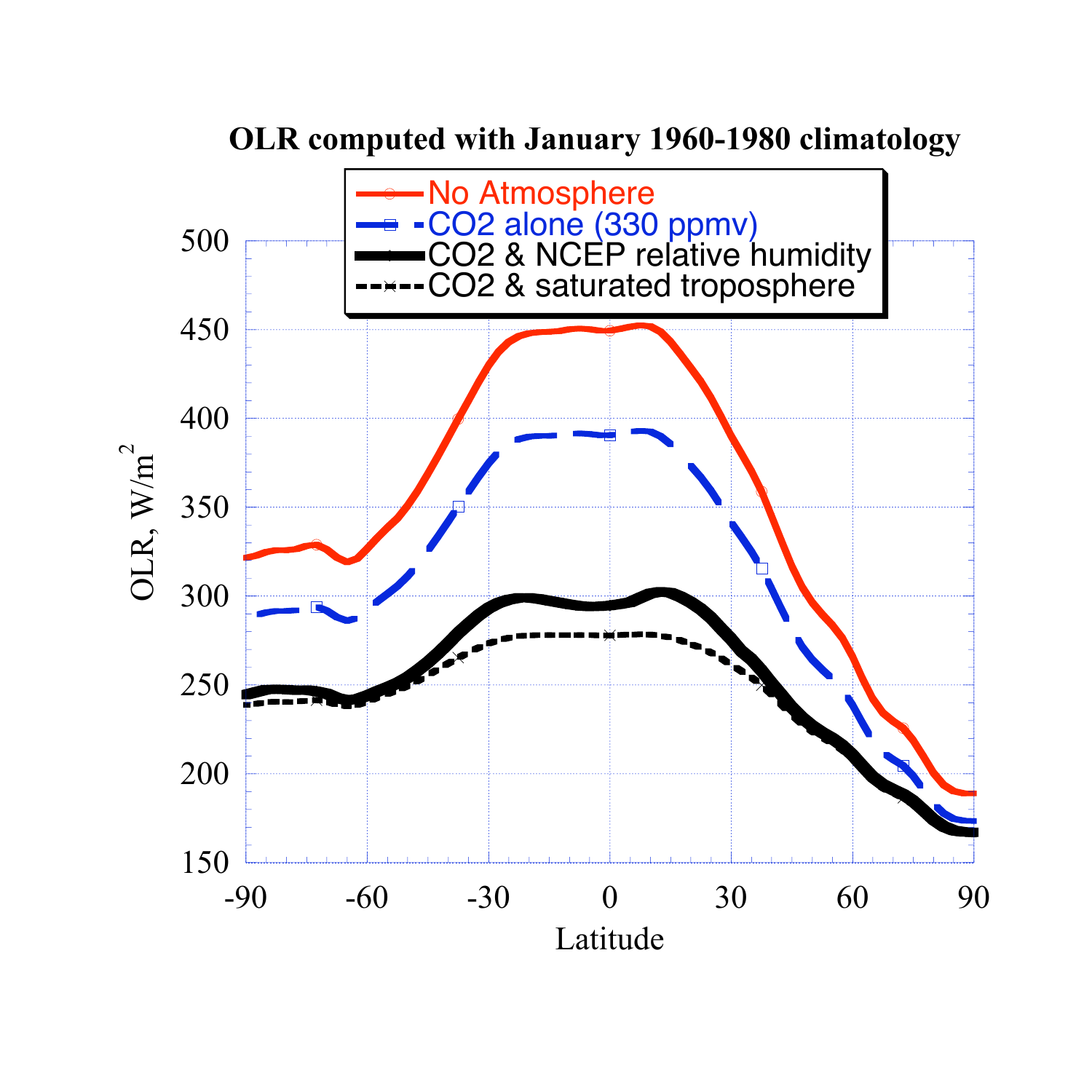, width=4in}
\caption{The effect of atmospheric composition on OLR,with atmospheric
and surface temperature held fixed at January climatological values.}
\label{fig:OLR}
\end{figure}

In this section we present a few reminders of the central role that
water vapor radiative effects play in determining the Earth's 
climate.  We begin by examining the way in which $CO_2$, water vapor,
and the presence of subsaturated air individually affect the radiation
balance of the Earth, as measured by $OLR$.  To do this, we used
the NCAR CCM3 radiation model \citep{CCM3} to compute what
the Earth's $OLR$ would be under various assumptions.  For each case clear-sky $OLR$
was computed on a latitude-longitude grid, with the 3D temperature pattern
held fixed at the January monthly mean climatology derived from NCEP
reanalysis data \citep{NCEP} for 1960-1980.  
The zonal mean results, shown in Figure \ref{fig:OLR}, are derived from
averages of latitude-dependent $OLR$, rather than being computed on the basis of
zonal mean temperature fields.  This is important, because radiation
is nonlinear in temperature and humidity. We have not, however, 
taken the effect of daily fluctuations into account.
This introduces some error into the analysis, particularly in the midlatitudes.

As compared to the calculation with no atmospheric greenhouse effect
whatsoever, $CO_2$ by itself brings the tropical $OLR$ down by
$60W/m^2$. The $OLR$ reduction decreases with latitude in the
extratropics, falling to $25 W/m^2$ at 60S in the summer hemisphere, and
even smaller values in the winter extratropics. The latitudinal
variation in the $CO_2$ greenhouse effect derives from the vertical
structure of the atmosphere: In the tropics there is more contrast
between surface and tropopause temperature than there is in the
extratropics, and the summer extratropics has more contrast than the
winter extratropics.  When $OLR$ is recalculated with the observed
humidity content of the atmosphere (based on NCEP) in addition to
the $CO_2$ from the previous case, $OLR$ drops by an additional
$100W/m^2$ in the tropics, and a lesser amount in the extratropics.
In fact, at each latitude the greenhouse effect of water vapor is
approximately twice that of $CO_2$.  As noted previously, the atmosphere
is highly unsaturated. If we recompute the $OLR$ assuming a saturated
troposphere, the $OLR$ is further reduced. The effect is considerable
in the tropics, amounting to over $20 W/m^2$ of additional greenhouse
effect.  In the summer midlatitudes, the effect of saturating the atmosphere
is much weaker, amounting to only $7 W/m^2$ at 45S.  This is because
the monthly mean relative humidity above the boundary layer is already
on the order of 50\% in the summer extratropics, so one is only doubling
the assumed moisture content there.  The winter extratropics shows a
somewhat weaker effect, on the order of $5 W/m^2$.  These results 
underestimate the effect of dry air on midlatitude $OLR$, because
the more extreme dry-moist contrast seen in daily data is associated
with mobile synoptic systems that average out in the monthly means.

\begin{figure}
\epsfig{file = 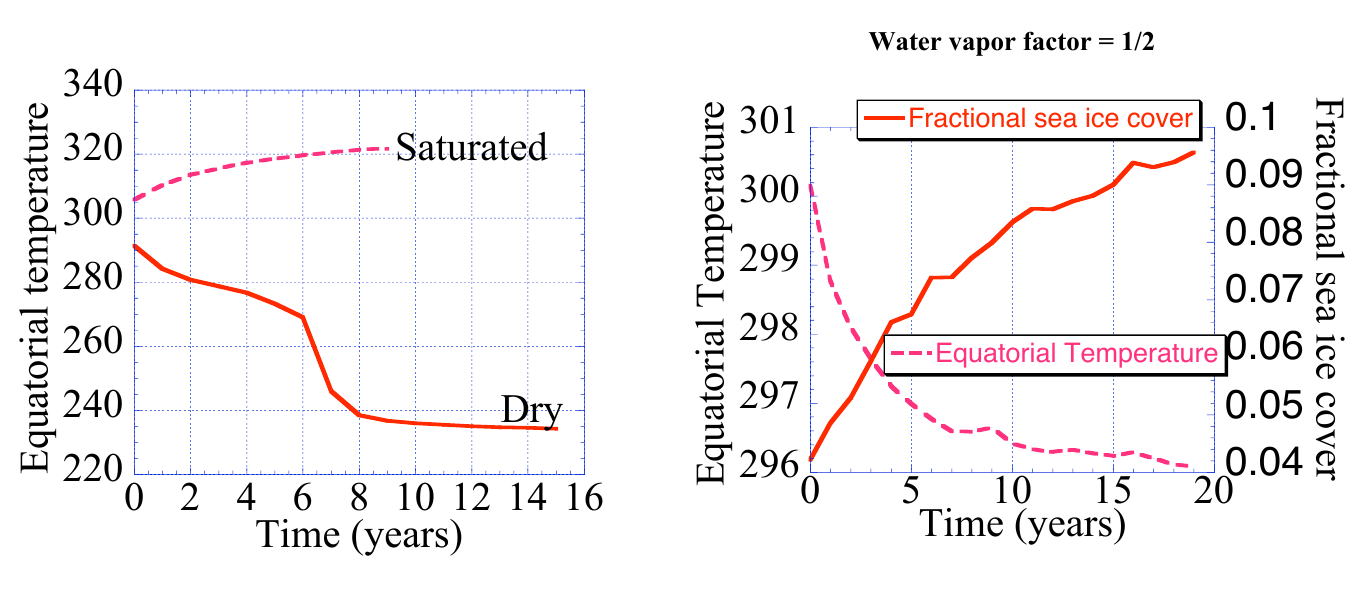,width=4.5in}
\caption{Results of GCM simulations in which the radiative impact of water vapor
has been artifically altered. Left panel:  Time evolution of equatorial temperature
for completely saturated and completely dry atmospheres.  Right panel: Time evolution
of equatorial temperature and global ice cover for a case in which the water vapor
used in computing radiation is reduced by a factor of two compared to the standard
parameterization.}
\label{fig:FOAMwater}
\end{figure}

Thus, a much drier atmosphere has the potential to be considerably colder
than the present climate, whereas a saturated atmosphere has the potential
to be considerably warmer.  To translate this statement into more quantitative
terms, we have carried out a series of GCM experiments in which the water vapor
content of the atmosphere has been artificially altered in the radiation
calculation alone. This approach allows one to study the radiative impacts
of water vapor in isolation, without needing to confront the formidable
task of disentangling them from the myriad other complex influences of
the hydrological cycle on climate.  It is essentially the same technique
employed by \citet{HallManabe99} in their study of water vapor feedback.
Our experiments were carried out with the FOAM GCM coupled to a mixed
layer ocean and thermodynamic sea ice model, employing realistic present-day geography.
\footnote{When coupled to a mixed layer ocean, the FOAM GCM is essentially
a portable, Beowulf-oriented re-implementation of CCM3 \citep{CCM3}.  All
simulations reported in this chapter were carried out at R15($4.5^o x 7.5^o$) resolution.
Further details on FOAM can be found at {\tt www.mcs.anl.gov/foam}.}
Each simulation started from a control run modern climate with
unaltered water vapor. The evolution of the climate following the
imposition of each altered water vapor assumption is shown in Figure \ref{fig:FOAMwater}.
When the radiative effect of water vapor is eliminated, the Earth falls
into a globally glaciated snowball state after 8 years.  To probe the opposite
extreme, we forced the radiation to be calculated under the assumption that
the entire troposphere is saturated.  Note that this does not mean that the
specific humidity was fixed at the value the control climate would have
had in saturation. Rather, each time the radiation module is called, the
saturation specific humidity is computed using the model's temperature field
at that instant of time; this specific humidity field is then used in the
radiation calculation.  This technique yields additional warming, because
the radiative forcing caused by saturating the atmosphere at any given
temperature is {\it itself} subject to amplification by water vapor feedback,
as it should be.  In the saturated experiment the tropical temperature
rises to an impressive $320K$ after 8 years.  It appears to have leveled
off at this time, but we do not know if the climate would continue to
warm if the integration were carried out for a longer time, since the 
simulation halted owing to numerical instabilities arising from extreme
temperature contrast between the Antarctic glacier and the surrounding
warm ocean waters. The tropical warming in the saturated GCM simulation
is similar to what we predicted earlier on the basis of an idealized
two-box model of the tropics \citep{ChapmanWater}.  The saturated case
shows the potential for water vapor feedbacks to make the climate response
to $CO_2$ much more extreme than found in the extant climate models; there is
plenty of dry air around, and if it somehow became saturated the consequences
for climate would be catastrophic.  The saturated case provides only an upper bound 
on how bad things could get, but it is nonetheless
an upper bound that should be kept in mind when thinking about climate prediction
uncertainties.  

We also examined the impact of a less extreme assumption about water vapor,
results for which are shown in the right panel of Figure \ref{fig:FOAMwater}.
In this case, the water vapor was computed using the GCM's standard
explicit and parameterized physics package, but the resulting specific humidity
was muliplied by one half before being handed to the radiation scheme.  
In this run, the equatorial temperature drops $4K$, to $296K$.  Moreover,
the fractional sea ice cover doubles, to nearly 10\%.  The summer snow line
on land (not shown) also moves considerably equatorward.  All in all, the
resulting climate rather resembles the Earth's climate during the Last Glacial
Maximum.

Removing the atmosphere's water provokes a Snowball Earth.  Saturating the
atmosphere provokes an extreme hothouse climate warmer than any encountered
in the past several hundred million years.   Reducing water
to half that produced by standard parameterizations is sufficient to
provoke an ice age.  Water vapor is a dynamic and fluctuating quantity
with very little intrinsic persistence beyond the monthly time scale.  
It is striking that, despite this fact, our climate doesn't undergo
massive fluctuations in response to spontaneously generated fluctuations
in water vapor content.  The lack of such fluctuations is a testament
to the precision with which water vapor statistics are slaved to
the large scale state of the climate. 

Water vapor is the atmosphere's single most important greenhouse gas,
and it is correct to say that an accurate prediction of climate
change hinges on the ability to accurately predict how the water
vapor content of the atmosphere will change. It would be a gross error,
however, to conclude that the effect of $CO_2$ is minor in comparison
to that of water vapor.  The greenhouse effect of $CO_2$ accounts for
fully a third of the total, and this is a very considerable number.
Eliminating the $50W/m^2$ of tropical $CO_2$ greenhouse effect would
drop the tropical temperature by about $25K$, once amplified by
water vapor feedback.  When further amplified by ice-albedo feedback,
this would certainly cause the Earth to fall into a snowball state.  
A warmer, water-rich world subjected to much greater solar radiation
than the Earth could indeed be in a state where the addition or removal
of a few hundred ppm of $CO_2$ would make little difference to the climate,
but this state of affairs does not by any stretch of the imagination apply
to the Earth. Another important difference between $CO_2$ and water
vapor is that the former has a much longer response time because, on Earth,
it is removed only by slow biogeochemical processes as opposed to
the rapid condensation and precipitation processes affecting water vapor.
Water vapor responds quickly to temperature changes induced by $CO_2$,
whereas $CO_2$ would take thousands of years to respond to a change
in the hydrological cycle.

\section{How saturated is the atmosphere?}
\label{sec:HowSaturated}

\begin{figure}
\epsfig{file=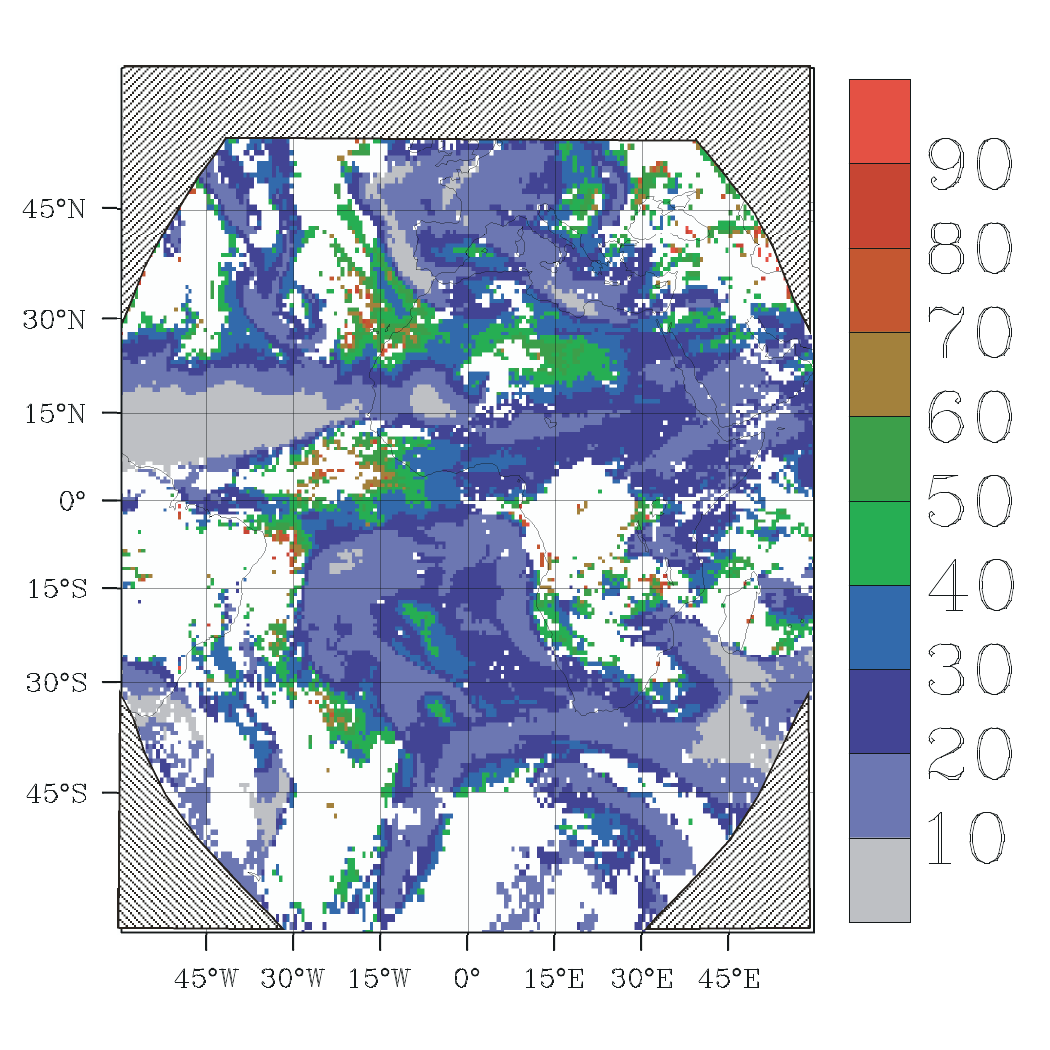,width=4in}
\caption{Free Tropospheric Relative Humidity on 16 January, 1992, retrieved from the Meteosat 
water vapor channel,based on retrieval in \cite{BrogniezThesis}. Cloudy regions where
the infrared retrieval could not be done were filled in with 100\% relative humidity for
the purposes of this figure. (See color version in online supplement for an indication
of the regions where cloud-clearing could not be done.)}
\label{fig:Meteosat}
\end{figure}

Above the boundary layer, much of the atmosphere is highly unsaturated,
with relative humidities below 10\% occurring frequently in both
the Tropics and the Extratropics.  The satellite snaphot of mid tropospheric
humidity on 16 Jan, 1992, shown in Figure \ref{fig:Meteosat},  serves as a reminder
of the prevalence of dry air.  The prevalence of dry air can be quantified
in terms of the probability distribution of relative humidity. This diagnostic
has been extensively studied in the Tropics \citep{spenbras97,ZhangBimodal,BrogniezThesis}.  
The midlatitude distribution has not received as much attention, but there
is ample evidence that highly undersaturated air is common there as
well (e.g. \citet{SodenBretherton}).  Early work suggested a lognormal
distribution of relative humidity \cite{SodenBretherton} but further
study has revealed a greater variety of shapes of the relative
humidity histogram, including even bimodality \citep{ZhangBimodal,BrogniezThesis}.  For now, the
detailed shape of the probability distribution will not concern us; it is
enough to know that most of the free troposphere is significantly undersaturated.
This widespread occurrence of unsaturated air is a manifestation of the fact that
from a thermodynamic standpoint, the atmosphere is a non-equilibrium system.
In a state of thermodynamic equilibrium with the oceanic moisture reservoir,
the entire atmosphere would be saturated.  The prevalence of subsaturated air, and
the sorting of air into moist and dry regions, yield a state of relatively low thermodynamic
entropy.  Evaporation of liquid water or ice into the subsaturated regions, or
diffusion of moisture from regions of high vapor pressure into 
subsaturated regions of lower vapor pressure, would increase the entropy of
the atmosphere \citep{PauluisHeld2002}.

%**GREYSCALE image substitute

It is essential to understand the processes that determine the
degree of subsaturation of the atmosphere, and to come to an understanding
of how the net result of these processes might change as climate changes.
In thinking about atmospheric water vapor content, one must draw a distinction
between the net column water content, and the water content in the mid to upper
troposphere. The latter accounts for a small fraction of the total water
content of the atmosphere, because the lower warmer parts of the atmosphere
have much higher saturation vapor pressure and are moreover sufficiently close
to the surface water reservoir that they tend to stay relatively saturated.
Yet, water vapor, or any greenhouse gas, placed near the surface has little
effect on OLR, because the low level air temperature is not much lower
than the surface temperature.  To get a significant greenhouse effect, one must
increase the infrared opacity of a portion of the atmosphere that is significantly
colder than the surface.  Because the radiative effect of water vapor is
logarithmic in its concentration, small quantities of water vapor can accomplish
this task aloft.  
This leads to the concept of "Free Tropospheric Humidity," (FTH) or
"Upper Tropospheric Humidity" (UTH) which may be loosely defined as the
water content of the portion of the atmosphere where water vapor has a
considerable effect on the radiation budget. Diversion of a tiny
proportion of the atmosphere's net water vapor content would be
sufficient to saturate the mid to upper troposphere and radically warm
the climate. For example, consider a 50mb thick layer of saturated air
near the surface of the tropical ocean, having a temperature of 295K.
Less than 3\% of the water content of this layer would suffice to
completely saturate a layer of equal mass having a temperature of
250K, such as would be encountered in the tropical mid-troposphere, or
at lower altitudes in the extratropics.  Clearly, the magnitude of the
boundary layer water vapor reservoir is not the limiting factor in
determining the free tropospheric humidity.    

We pause now to dismiss two fallacies, which unfortunately have not yet
completely disappeared from discussions of the subject at hand.  The
first is the {\it evaporation fallacy}, typified by statements like: "In
a warmer world there will be more evaporation, which will increase the
supply of moisture to the atmosphere and increase its humidity."  This
is wrong on at least two counts.  It is not inevitable that warmth
increases evaporation, since evaporation is determined by wind speed and
boundary layer relative humidity as well as the boundary layer
saturation vapor pressure (which does unquestionably go up with
temperature). It is entirely possible for a warmer climate to have less
evaporation.  An even more serious objection to the evaporation fallacy
is that it hopelessly confuses fluxes and reservoirs.  Evaporation gives
the {\it rate} at which moisture fluxes through the atmosphere, which is
distinct from, and even has different units from, the amount of water
left behind in the atmosphere (or more specifically the free
troposphere). In this regard, the evaporation fallacy is as absurd as
saying that, in comparing the water content of two vessels, the one with
the higher rate of filling will contain more water, regardless of
whether one of them is a sieve while the other is a watertight bucket. 
The second fallacy is the {\it saturation fallacy}, which states flatly
that a warmer atmosphere will tend to become moister because saturation
vapor pressure increases with temperature. That is like saying that one
always expects to find more water in a bigger bucket than a smaller
bucket, no matter how leaky either might be. Clausius-Clapeyron does
indeed govern the relation between temperature and water content, but in
a far subtler way which will begin to become clear shortly.

So what really determines the water vapor content of the free
troposphere? It is easiest to think about this problem in a Lagrangian
sense, tracking the water content of an air parcel as it wanders about
the atmosphere. The fluctuating water content of the parcel results from
a balance between the rate at which water is added to the parcel against
the rate at which water is removed.  Water vapor is removed either by
condensation or by diffusion into a neighboring drier air parcel. Let us
suppose for the moment that diffusivity is so low that the latter
mechanism is unimportant.  In that case, water vapor is removed when the
air parcel wanders into a region where the local saturation specific humidity
is lower than the current specific humidity of the parcel,  at which time the
specific humidity is reset to the lower local saturation value and the
balance is rained out. The net result is that the specific humidity of an
initially saturated parcel after time $\tau$ is equal to the minimum
$q_s$ encountered along the trajectory during that time. By definition,
this is a non-increasing function of $\tau$, though there will be long
periods of time over which the minimum remains constant, between
those times at which new minima are encountered. Define
$q_{min}(t_o,\tau) \equiv \min(q_s(t);t_o < t < t_o+\tau)$. Then,
the rate at which $q_{min}$ decreases, which is different for each trajectory,
determines the drying rate on that trajectory.  The drying is not
a continual process, but occurs in fits and starts as new minima
are encountered. If no new water is added to the parcel,
new condensation events eventually become very infrequent,
or cease altogether if the global minimum is encountered.  
The drying rate determined by this process must be balanced against
whatever processes episodically add new water to the air parcel,
determining the statistically fluctuating water content of the parcel. 

The importance of nonlocal control of tropospheric humidity was clearly 
revealed in the case study described in \citet{KTD91}. Drawing on this insight,
\citet{YP94} examined the general problem of midlatitude dry air production from
a Lagrangian standpoint, and pointed out the importance of the $q_{min}$ statistic.
The most prominent dry air pool in the atmosphere is found in the subtropics,
and over the next several years a number of investigators developed an interest
in large scale control of subtropical humidity, employing Lagrangian
techniques \citep{EP96,salhart97,soden98,P98,PR98,SherwoodDessler2000,Galewsky2005}.
It was an idea whose time had come, and these studies provided a reassuring
counterpoint to the then-emerging suspicion that everything was controlled
by parameterized microphysics, and everything was inscrutable. 
In recent years, the term {\it time of last saturation model} has gained currency as a name
for the general approach. We prefer the broader term {\it advection-condensation model}.
If one knows the specific humidity of the air parcel at its destination, then one can
determine the time of last saturation by following the back-trajectory until that
specific humidity becomes saturated.  However, if the aim is to {\it predict} the
specific humidity at the destination,  determining the time of last saturation requires
some assumption about the processes which, at some time in the past, added moisture
to the air parcel.

\begin{figure}
\epsfig{file = 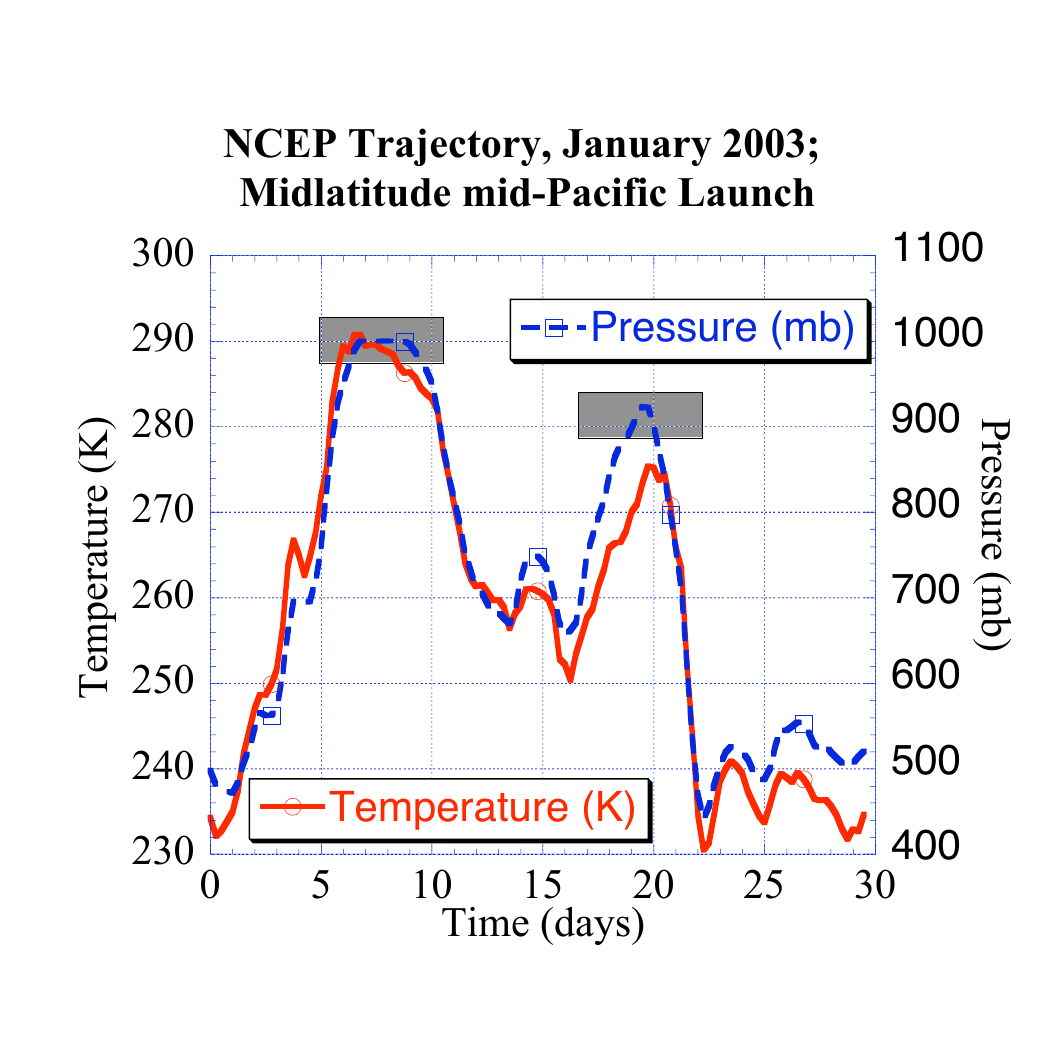 , width = 4in}
\caption{Pressure and temperature along a 3D air parcel trajectory
launched at 500mb in the midlatitude Pacific.  The trajectory was computed
on the basis of 4x daily NCEP horizontal and vertical velocity fields. The
grey boxes indicate encounters with the boundary layer, where moistening
is presumed to occur.}
\label{fig:ncepTrajec}
\end{figure}

Figure \ref{fig:ncepTrajec} shows temperature and pressure following a typical
midlatitude trajectory.  The temperature and pressure are nearly perfectly
correlated, so minima in $q_s$ are very nearly coincident with minima
in $T$ or $p$.  The trajectory executes a wavelike motion under the influence
of midlatitude synoptic eddies, swinging between the subtropical surface
where the temperature reaches 290K, and the high latitude tropopause, where
the temperature drops to 230K.  If there were no moisture sources in the free
troposphere, then we would know that when the trajectory reaches the 800mb
level on day 5, its specific humidity would be no greater than $1.4\cdot 10^{-4}$, the value
corresponding to the minimum at day 0 (computed with 232K and 500mb). If the
parcel becomes saturated when it encounters the ground at day 7, then it 
remains at saturation through day 13 since the temperature and saturation
specific humidity decrease monotonically through that part of the trajectory. However, when the
parcel lands on the 800mb surface on day 18, its specific humidity would
be $6.6\cdot 10^{-4}$, corresponding to the minimum at day 16.  Based on
the local saturation specific humidity on day 18, the relative humidity would
be 25\%. This trajectory illustrates an important principle:  in order
to know the specific humidity at a given point on the trajectory, one only needs
to know the history going back to the time of last saturation.  Information
about earlier times is erased whenever saturation occurs.  The trajectory
also illustrates the fact that midlatitude trajectories typically dip into
the subtropical boundary layer every 15 days or so, where they have an opportunity
to pick up new moisture. This is probably the main moisture source for the
midlatitude free troposphere.

Most midlatitude trajectories have a character qualitatively like that
shown in Figure \ref{fig:ncepTrajec}, but tropical trajectories are
quite different.  Isentropes are flat there, and there is less
baroclinic eddy activity.  Trajectories are dominated by organized
rising motion in convective regions, ejection due to steady outflow or
tropical transient eddies, and slow descent in the large scale organized
subsidence regions (see \citet{P98,PR98}).  The tropical model of
\citet{MD2004} provides perhaps the simplest realization of this class of
models, in that the trajectories consist simply of ejection of saturated
air from convective regions, followed by monotone subsidence at a rate
determined by infrared radiative cooling. This idealized trajectory
model neglects the lateral mixing incorporated in \citet{PR98}, but
nonetheless yields a number of insights concerning the nature of water
vapor feedback. Generally speaking, one can expect the statistics
characterizing trajectories to differ considerably between the tropics
and midlatitudes.

If we consider an ensemble of trajectories launched from a given
location, the behavior of the minimum $q_s$ statistic can be
characterized by the probability density $Q_{min}(q_{min}|q_o,\tau,t)$,
which gives the probability that the minimum $q_s$ encountered between
time $t$ and $t+\tau$ is near $q_{min}$, given that $q_s = q_o$ at the
place where the particle is located at time $t$. Obviously $Q_{min} = 0$
for $q_{min} > q_o$.  If the process is statistically stationary,
$Q_{min}$ will be independent of $t$.  If one is trying to understand
the water vapor at a specified point, it is most convenient to deal with
back trajectories, corresponding to negative $\tau$.  We are more
interested in where the air arriving at the target point {\it came from}
than in where it is going. If the trajectory process is statistically
reversible, all statistics of back trajectories have the same behavior
as the corresponding statistics of forward trajectories, and in
particular $Q_{min}(q_{min}|q_o,\tau,t) = Q_{min}(q_{min}|q_o,-\tau,t)$.
Somewhat counter-intuitively, Brownian motion is statistically
reversible in this sense.  It is widely assumed, and probably true, that
the atmospheric trajectory problem is statistically reversible, though
we will not explicitly make use of the assumption in our calculations
below.

As an example of the use of back-trajectory statistics, let's suppose
that we wish to know the probability distribution of specific humidity $q$
in some patch of the 500mb surface at a given time $t$, and are willing
to assume that the entire atmosphere was saturated at time $t-\tau$, and
moreover that there were no moisture sources in the intervening time.
To solve this problem, we use $Q_{min}$ for the ensemble of trajectories
which arrive at the patch at the specified time.  For simplicity, we
will assume that $q_o$ is constant within the patch, though this is an
assumption that can be easily relaxed.  Because we assumed that all
parcels are saturated at time $t-\tau$ (though each has a different
$q_s$, appropriate to its location at that time), the specific humidity
each parcel winds up with by time $t$ is simply the minimum $q_s$ 
encountered along the trajectory. Hence, the probability density function ("PDF")
of $q$ in the patch is $Q_{min}(q|q_o,-\tau,t)$, which is concentrated
on progressively drier values as $\tau$ is made larger.  Note that
this distribution is entirely distinct from the distribution of 
{\it initial} saturations at time $t-\tau$.  These could
all be in the tropical boundary layer and have very high values, and the
final humidity at time $t$ would still become small.  A {\it passive} tracer,
with no sources or sinks, would retain its initial value, so that its 
PDF at later times is determined solely by its initial PDF, with no knowledge
of the nature of the intervening path required.  In contrast, the statistics
of moisture are sensitive to the entire history of the path taken.  The
sensitivity to probabilities that depend on entire paths is one of the
chief mathematical novelties of the water vapor problem, and the source
of most of the theoretical challenges.

$Q_{min}$ characterizes the drying process, and one needs a corresponding
probabilistic description of the moisture source in order to complete a
theory of the atmospheric humidity.  A moisture source such as evaporation
of precipitation falling through dry air could add just a bit of moisture
to an air parcel without saturating it. However, let's idealize the moisture
source as a series of saturation events, which occur randomly in time,
with the chance that a saturation event will occur after waiting
for a time $\tau$ described by a probability distribution
$P_{sat}(\tau)$.  Then, the PDF of specific humidity is given by
the convolution 
\begin{equation}
Q(q|q_o,t) =  \int_0^\infty P_{sat}(\tau)Q_{min}(q|q_o,-\tau,t) d\tau
\label{eqn:MoistureConvolution}
\end{equation}
where $q_o$ is the saturation specific humidity at the point under
consideration. If the trajectory process is statistically stationary,
$Q$ will be independent of $t$. 

As an example of the application of Eq. \ref{eqn:MoistureConvolution} in the
simplest possible context, we consider a uniform subsidence model similar to
that in \cite{MD2004}.  In this case, since the trajectories always go downwards
to regions of larger $q_s$, the minimum $q_s$ in the back-trajectory always
occurs at the time of saturation.  To be definite, let's assume that
the vertical coordinate is pressure $p$, that trajectories subside at
a constant rate $\omega$, and that $q_s = q_s(p)$, i.e. that the saturation
field is horizontally homogeneous, as is approximately the case in the tropics.
Then, $Q_{min}$ is a $\delta$-function concentrated on the value of $q_s$
at the pressure the parcel was at when it was resaturated. If $p_o$ is
the pressure at the target point (i.e. $q_o = q(p_o)$) then the pressure
at the saturation point is $p_o - \omega \tau$ and hence
$Q_{min}(q|q_o,-\tau,t) = \delta(q - q_s(p_o - \omega \tau)$. Substituting
this into Eq. \ref{eqn:MoistureConvolution} and using the relation
$\delta(g(\tau))d\tau = (g'(\tau))^{-1}\delta(g) dg$ we find
\begin{equation}
Q(q|q_o,t) = \frac{1}{\omega \frac{dq_s}{dp}|_{p_o - \omega \tau^*}} P_{sat}(\tau^*)
\label{eqn:SubsidencePDF}
\end{equation}
where $\tau^*(q)$ is the solution to $q_s(p_o - \omega \tau^*) = q$. In this
case, the specific humidity PDF is determined by the saturation statistics
and the vertical structure of $q_s$  
\footnote{ The recent work of Sherwood and Meyer (personal communication)
on relative humidity PDF's employs a special case of this class of solutions}.
In the general case, the moisture
dynamics is characterized by the two PDF's $P_{sat}$ and $Q_{min}$.  In
order to fully understand water vapor feedbacks, we need to understand
how these two PDFs change as climate changes. This is a tall order. Some
further examples of the trajectory statistics in action will be given in Section
\ref{sec:SimpleModels} for idealized trajectory models, and Section \ref{sec:RealTrajec} for
realistic trajectories.

There are three difficulties with the trajectory approach, two of them
technical, and one of a more fundamental nature.  The first difficulty is that there may
be mixing of moisture amongst air parcels arising from small scale
turbulent motions.  Because large scale resolved strain causes
exponential amplification of gradients \citep{YP94,P98} even a weak effective
diffusivity would eventually cause significant mixing.  The mixing
is likely to be dominated by vertical rather than horizontal mixing
processes, for the reasons discussed in \citet{HaynesAnglade}.  Incorporation
of mixing greatly complicates the calculation, because the moisture evolution
on one trajectory becomes dependent on the moisture evolution of all other
trajectories which pass in its vicinity during the past. Explicit calculation
then calls for either Eulerian methods (at the expense of the need to
confront the difficult problem of unwanted numerical diffusion), or 
simultaneously integrating the problem on a sufficiently large swarm of
trajectories.  Mixing between moist and dry air parcels is important
because it both reduces the frequency of very dry air, and because 
the dilution increases the subsaturation of the moist air, and delays 
condensation.  It is of utmost importance to determine how
much small scale mixing enters into the tropospheric moisture problem
in the real atmosphere, and comparison of observations with results
of the non-diffusive trajectory calculation can provide a means
of doing so.  There has been considerable attention to the diagnosis
of small scale mixing in the stratosphere (see \citet{HaynesAnglade,Legras2003} and
references therein), but the analogous question for tropospheric water vapor
is unresolved.  The second difficulty with the trajectory method is
similar to the mixing problem:  precipitation generated by large scale
condensation along a trajectory may evaporate as it falls through dry
air, adding moisture to trajectories it encounters. This, too couples
trajectories.  It can be regarded as just another form of vertical
mixing.

The third difficulty comes at the cloud scale, where we confront a much
harder problem, particularly in tropical convective regions.  Models and
analyses produce a large scale rising motion in such regions,
diagnostically associated with the moisture convergence required to feed
the convective precipitation. This upward motion, which lifts all
trajectories, must be regarded as wholly fictitious.  In reality,
most of the air in the convective region is subsiding, and the
upward mass flux is concentrated in cumulus towers covering only a small
fraction of the area of the convective region.  The resolved large-scale
vertical velocity correctly represents the net upward mass flux, but is
not typical of the velocity of individual fluid parcels in the
convective region. The net result is that the detrainment obtained from
large scale trajectory simulations is always saturated, whereas the
cloud scale motions offer ample opportunities for the detrained air to
be substantially undersaturated. The mixing of moisture between moist
and dry air at cloud and subcloud scales engages microphysical issues of
the sort discussed by \citet{TompkinsEmanuel2000}, and ultimately
determines the degree of subsaturation. The success of models with
saturated detrainment at reproducing water vapor observed in
nonconvective regions (e.g. \citet{MD2004,PR98}) suggests that the
subsaturation must not be too extreme in the typical case, but the whole
matter requires further study.

\section{A few illustrative models}
\label{sec:SimpleModels}

It is instructive to consider some simple one-dimensional models of the
connection between mixing, transport and drying. These models are
offered in order to highlight some of the basic issues in cases which
can be solved completely. We do not pretend that any of these models
yield good representations of the atmosphere's actual water vapor
distribution. Both of the models we consider in this section are
formulated in terms of an abstract spatial coordinate $y$.  The only
thing we need to know about $y$ is the saturation specific humidity $q_s$ as
a function of $y$. We may think of $y$ as the North-South distance
following an isentropic surface, which yields a decreasing $q_s$ because
the surface becomes generally higher (hence colder) as the pole is
approached.  The coordinate could represent North-South distance at a
fixed mid-tropospheric pressure level, as in typical one layer energy
balance models.  With a suitable increase in mixing rates, $y$ could
equally well represent altitude, with latitude and longitude held fixed;
in this case the mixing parameterization is to be thought of as a
surrogate for convection rather than the larger scale, more ponderous,
large scale advection.  If temperature decreases linearly with
$y$, then the Clausius-Clapeyron relation implies that $q_s$ decreases
approximately exponentially with $y$.  A decrease of pressure
with $y$ somewhat offsets the exponential decay, but not so much
so as to prevent us from using $q_s(y) = \exp(-y)$ as a useful
conceptual model of the atmosphere's saturation specific humidity.

\subsection{The Diffusion-Condensation model}
\label{sec:DiffCon}
We wish to study the interplay of transport and condensation. Diffusion
is the simplest model of transport, and is moreover employed
as a surrogate for eddy transport of water vapor in many idealized
climate models (e.g. \citet{UVicModel,Climber}).  Hence, 
we first examine a model in which moisture is represented by the mean specific humidity
$q(y,t)$, which is stipulated to satisfy a diffusion equation

\begin{equation}
\partial_t q - D \partial_{yy} q = -S(q,q_s)
\label{eqn:diffcon}
\end{equation}

In this equation, the sink $S(q)$ instantaneously resets $q$ to $q_s$ whenever diffusion
causes it to exceed $q_s(y,t)$. The instantaneous sink may be taken as a
limiting form of the function
\begin{equation}
S(q,q_s) = 
  \begin{cases}
       \frac{1}{\tau} (q - q_s(y,t)) &\text{for $q > q_s$} \\
          0 &\text{for $q \le q_s$}
  \end{cases}
\end{equation}
as $\tau \rightarrow 0$.  The mathematical novelty that a sink of this
form adds to the problem is that it makes the problem nonlinear.  Condensation
can be thought of as a particularly simple form of unary chemical reaction,
and in fact many of the issues we encounter in the condensation problem are
generic to a broad class of nonlinear chemical reactions.

In general, $q_s$ could be a function of both $y$ and $t$, but henceforth
we shall assume $q_s = q_s(y)$. In the instantaneous removal limit
($\tau \rightarrow 0$), $q(y) = q_s(y)$ is a steady solution to Eq. \ref{eqn:diffcon}
in any region where $d^2q_s/dy^2 > 0$.  If this condition is satisfied at
time $t$, then $S=0$ at that time but the diffusion term makes $\partial_t q > 0$,
so that the moisture sink will be activated at the next instant of time, and reset
the moisture to $q_s$, keeping the moisture in the region fixed at $q_s(y)$
\footnote{ In the regularized case with small but finite $\tau$, the removal process is 
less singular and the steady solution has $q$ very slightly greater than $q_s$ and
nonzero removal rate $S$.}.
The moisture loss is balanced by moisture flux from regions of larger $q$.

Next, we consider a simple initial value problem. Suppose that $y$
extends from $0$ to $\infty$,  and impose the no-flux boundary condition
$\partial_y q = 0$ at $y=0$ so that there are no sources of new
moisture.  In this case, the condensation caused by diffusion of
moisture from regions of large $q_s$ into regions of small $q_s$ will
deplete the moisture content. We shall assume that  $d^2q_s/dy^2 > 0$
throughout the domain, and that $q_s$ attains its maximum value at
$y=0$. Given the curvature assumption, this implies that $q_s$ is
monotonically decreasing. If the atmosphere is initially saturated, then
there is some $Y(t)$ such that $q(y) = q_s(y)$ for $y > Y(t)$; this is
the region in which condensation is taking place.  For $y < Y(t)$ the
air is undersaturated, and satisfies the diffusion equation with no
sources or sinks.   As moisture is drawn out of the system by
condensation, $Y(t)$ increases.  Under the assumptions on $q_s$, this
quantity approaches a minimum value $C$ from above as $y \rightarrow \infty$.  
As $t \rightarrow \infty$, we have $Y(t) \rightarrow \infty$
and $q \rightarrow C$. The question of the long time asymptotic
behavior of $Y$ is thus well posed, and in fact has a fairly simple and
general answer. Since the advection-condensation equation is invariant
under the transformation $q \mapsto q - C, q_s \mapsto q_s - C$, we can assume
that $C = 0$, without any loss of generality.

Examination of some numerical solutions suggests that $q$ is approximately
parabolic in the noncondensing region after a sufficiently long time
has passed.  Let us then look for a solution with $q = a - b y^2$ for $y < Y(t)$.
What we shall present here is not a complete and rigorous asymptotic analysis,
for we shall not derive the conditions under which the assumed form of $q$ in
the noncondensing region is valid. 
The matching conditions at $Y$ and the requirement that $q$ satisfy
the diffusion equation in the subsaturated region imply
\begin{equation}
 \begin{split}
    a - bY^2 = q_s(Y)\\
    -2bY = q_s'(Y)\\
    \frac{da}{dt} = -2bD
\end{split}
\end{equation}
Given the assumed form of $q$, we can only exactly satisfy the diffusion equation
at one point, and in writing the third of these relations we have chosen to do
so at $y=0$.
Taking the time derivative of the first equation and substituting from
the other two results in the following differential equation for $Y$:
\begin{equation}
-2bD + \frac{1}{2}q_s'(Y)  \frac{dY}{dt}  + \frac{1}{2}q_s''(Y) Y \frac{dY}{dt}
 = q_s'(Y)\frac{dY}{dt} 
\end{equation}
or, equivalently,
\begin{equation}
Y\cdot (1 - Y \frac{q_s''(Y)}{q_s'(Y)} )\frac{dY}{dt} = 2D
\label{eqn:AsympY}
\end{equation}
The quantity $q_s''/q_s'$ has dimensions of inverse length.  For $q_s = \exp(-y^2/\sigma^2)$
it is $-2Y/\sigma^2$ at large $Y$, for $q_s = \exp(-y/L)$ it has the constant value
$-1/L$ and for $q_s = A/y^\alpha$ it is $-(\alpha+1)/Y$.   Suppose
that $q_s''/q_s' = AY^n$ at large $Y$.  Then, if $n \le -1$ the second term multiplying
the derivative on the left hand side of Eq. \ref{eqn:AsympY} is at most
order unity, and the equation can be integrated to yield 
$Y(t) \sim \sqrt(Dt)$.  On the other hand, if $n > -1$ the second term dominates
and we find $Y \sim (Dt)^{1/(n+3)}$.  It is interesting that rapid decay
of $q_s$ can cause the thickness of the subsaturated region to grow 
more slowly than the diffusive length scale, $\sqrt(Dt)$.

Once $Y(t)$ is known, the humidity decay curve is obtained using
\begin{equation}
q(0) = a = q_s(Y) -\frac{1}{2} Yq_s'(Y) = (1- \frac{1}{2} Yq_s'(Y)/q_s(Y))q_s(Y)
\end{equation}
At large $Y$, $q_s'/q_s$ will generally have the same scaling as $q_s''/q_s'$; by
assumption, it is always negative.  If $q_s$ decays exponentially then
$q_s'/q_s$ is a negative constant and the second term in the factor multiplying
$q_s$ dominates; in this case $q(0) \sim Y q_s(Y)$, and hence decays like
$t^{1/3}\exp(-c t^{1/3})$ for some constant $c$.  If $q_s$ decays
algebraically, then both terms in the factor are order unity, so 
$q(0) \sim q_s(Y)$ and hence decays like $ t^{-\alpha/2}$, assuming
$q_s$ decays like $y^{-\alpha}$ at large $y$. In either case, it
can easily be shown by integrating $q(y)$ that the total
water vapor remaining in the atmosphere decays like $Y(t)q(0,t)$ provided
$q_s$ has a finite integral.
%Reasoning: Net water content is approx
%Y(t) (a + q_s(Y))/2, and a dominates or is comparable to q_s,
%so basically total water scales like Y(t)a(t)

\begin{figure}
\epsfig{file = 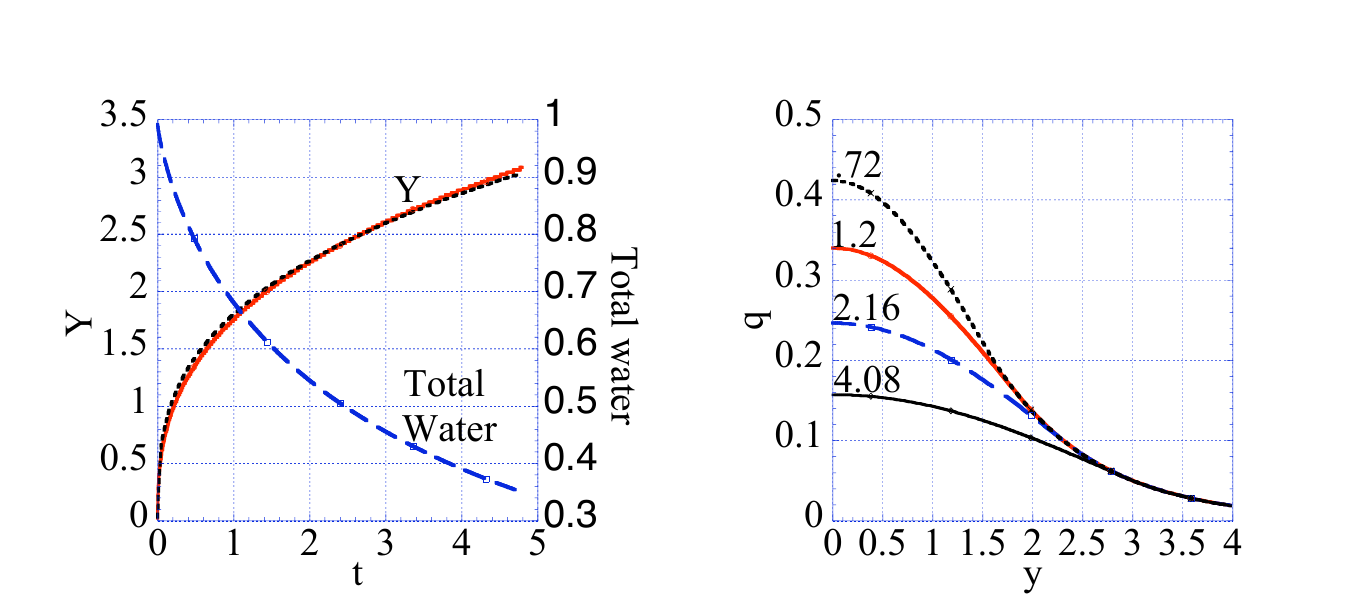,width=4.5in}
\caption{Numerical results for the freely decaying diffusion-condensation
model with a no-flux barrier at $y=0$.  Left panel: Time evolution of
the point $Y(t)$ bounding the subsaturated region, and of total moisture
in the system. The short-dashed line gives the fit to the asymptotic
result $Y \sim t^{1/3}$. Right panel:  The profile of specific humidity at
the times indicated on the curves.}
\label{fig:DiffConDecay}
\end{figure}

Thus, moisture is drawn out of the system by diffusion from subsaturated
regions neighboring the boundary ("the ground, in many cases") into regions of saturated air with
lower specific humidity at larger $y$, where it can condense. The rate of decay
is determined by the rate at which $q_s$ decays with increasing $y$.
Even for exponentially decaying $q_s$, the moisture decay is slower than
exponential, owing to the time required for moisture to diffuse into
the ever-retreating condensing region.  Figure \ref{fig:DiffConDecay}
shows numerical solutions to the problem with $q_s = \exp(-y)$, 
together with the theoretical result for $Y(t)$.  The agreement
between simulation and theory for this quantity is excellent.

As the next stage in our exploration of the diffusion-condensation equation,
we exhibit the steady state response to a moisture source.  
We introduce the moisture source, by holding $q = rh\cdot q_s(0)$ at $y = 0$ (with
$rh<1$).  If we assume $d^2q_s/dy^2 > 0$ as before, the equilibrium
moisture distribution has the simple form
\begin{equation}
q =
 \begin{cases}
   q_s(y) &\text{for $y > y_s$}\\  
    rh\cdot q_s(0) + \frac{(q_s(y_s)-rh\cdot q_s(0))y}{y_s} &\text{for $y < y_s$}
 \end{cases}
\end{equation}
where $y_s$ is chosen to make the flux continuous, i.e.
\begin{equation}
\frac{dq_s}{dy}(y_s) = \frac{(q_s(y_s)-rh\cdot q_s(0))}{y_s}
\end{equation}
All condensation is at $y > y_s$, and $y_s  \rightarrow 0$ as $rh
\rightarrow 1$, in which case the whole atmosphere becomes saturated. 
Further, although the undersaturation for $y < y_s$ relies on mixing,
$q(y)$ is independent of the magnitude of $D$. However, the rate of
moisture flux into the condensation region, and hence the precipitation
rate, is proportional to $D$. This solution underscores the point we made
earlier with regard to the "evaporation fallacy": The factors governing the
atmospheric relative humidity distribution are quite distinct from those governing
the rate at which water fluxes through the system. Note also that the moisture sink produces air
that has a lower specific humidity than the source, even in regions where no
condensation is taking place.  In effect, this is due to dilution of
moist air with drier air from larger $y$.  Still, the diffusive model
cannot produce an undersaturated layer unless the source air at $y=0$ is
undersaturated.

Another interesting configuration is the {\it cold trap}, in which
$q_s(y) = q_c < 1$ in a small region near $y=0$, while $q_s(y) = 1$ elsewhere.
A local minimum of this sort could be taken as a crude representation of
the minimum occurring at the tropopause, with $y$ measuring altitude.
However, the cold trap is also relevant to the case in which
$y$ represents horizontal distance on a fairly level surface, such as the 
Tropical Tropopaus Level (TTL); in this case, the
minimum corresponds to the cold conditions occurring over the high 
tropopause region above the Western Pacific Warm Pool, and we seek to
understand the global drying effect of this region. In the freely decaying
case subject to initial condition $q(y)=1$ in an unbounded domain, the cold
trap creates a zone around itself in which $q$ approaches $q_c$. The width
of this dry zone increases in proportion to $\sqrt{Dt}$.  Alternately,
we can seek an equilibrium solution by imposing $q = 1$ at $y = -1$ and
a no flux condition $\partial_y q = 0$ at $y = 1$.  The steady solution
in this case is simply $q = 1 -(1-q_c)(y+1)$ for $-1 \le y \le 0$ and
$q = q_c$ for $y \ge 0$.  In accord with intuition, moisture is reset
to the cold trap specific humidity when air passes through it.  We can 
also impose a moisture source at the right hand boundary, by replacing
the no-flux condition by $q = q_1$ at $y=1$.  In this case the solution is
\begin{equation}
q = 
 \begin{cases}
   1 -(1-q_c)(y+1) &\text{for $-1 \le y \le 0$}\\  
   q_c + (q_1-q_c)y &\text{for $0 \le y \le 1$}
 \end{cases}
\label{eqn:ColdTrap}
\end{equation}
{\it provided} that $q_c \le \frac{1}{2}(1+q_1)$. If this condition is met, the
solution takes the form of a bent stick, with the cold trap depressing the value
at the center of the domain. However, if $q_c > \frac{1}{2}(1+q_1)$, diffusive
dilution of the moist air reduces the humidity at the midpoint of the domain
to such an extent that no condensation occurs; in this case, the solution
is simply a straight line linking the limiting values at the two boundaries,
and is a solution of the conventional diffusion equation.  In Section 
\ref{sec:StochasticEquilibrium} we will treat the same configuration
using a stochastic model of water vapor, and find some intriguing differences
in the behavior.

%We could give the actual time-dependent solution in terms of error functions
%Future work:  Results with random q_s  

There can be no sustained condensation in a region where $d^2q_s/dy^2 < 0$.
In such a region, if $q = q_s$ initially, then diffusion will immediately reduce $q$ to
below its saturation value everywhere, halting condensation.  For example,
suppose $q_s = \exp(-y^2)$ in an infinite domain, and that $q=q_s$ initially.
After the initial instant, condensation halts for $|y|<1$, though it continues
for $|y|>1$ where the curvature is positive.  

The summary behavior of the 1D diffusion-condensation problem is quite
simple.  It creates saturated regions embedded in regions where
$q_s''$ is positive and $q_s$ is small. These drain moisture out
of the surrounding air, creating subsaturated regions in the vicinity.  
Apart from the possible inadequacies of diffusion as a representation of
the mixing, the main shortcoming of the diffusion-condensation model is that it
represents the moisture field in terms of a single concentration $q$ at
each $y$. In reality, a small box drawn about $y$ will contain an
intermingling of moist and dry air.  The diffusion model is
incapable of predicting a probability distribution function  for
moisture; worse, the neglect of fluctuations
fundamentally misrepresents the drying process
itself, owing to the nonlinear nature of condensation. The next
class of models we shall study rectifies this shortcoming of
the diffusion-condensation model.

\subsection{Stochastic drying: Initial value problem}
As a counterpoint to the diffusion-condensation problem we pose
a simplified random walk version of the stochastic drying
process introduced in Section \ref{sec:HowSaturated}.  Suppose
that an ensemble of particles execute independent random walks
in an unbounded 1D domain with coordinate $y$.  Particle $j$
is located at point $y_j(t)$, and is tagged with a
specific humidity $q_j$, which can change with time if condensation
occurs.  The saturation specific humidity field $q_s(y)$ is assumed
to be monotonically decreasing with y.  The particles are initially
saturated (i.e. $q_j(0) = q_s(y_j(0))$), but whenever they find
themselves at a place where $q_j > q_s(y_j)$, then $q_j$ is instantaneously
reset to the local $q_s$.  Under these conditions, what is
the probability distribution of specific humidity for the particles located
at point $y$ at time $t$?  

This problem is readily solved in terms of the maximum excursion
statistics for Brownian motion (i.e. a random walk with velocity
$\delta$-correlated in time). We make use of the fact that the
statistics for forward trajectories are identical to those
for backward trajectories, and pose the question as follows:
for a particle which is located at $y_b$ at time $t=0$, what is
the probability that it was located at $y_a$ at time $t = -\tau$?
What is the probability that the maximum $y$ visited in
the time interval $[-\tau,0]$ was $y_{max}$? Since $q_s$ is
monotononically decreasing, the latter probability gives
us the probability of $q_{min}$.  If we define
the random walk as satisfying $dy/dt = v(t)$ with
$<v(t)v(t')> = 2D \delta(t-t')$, then the   
probability that the parcel landing at $y_b$ came
from $y_a$ is given by the familiar Gaussian form, just
as if we were running the trajectory forwards rather
than backwards \citep{KS91,PileOuFace}:
\begin{equation}
p(y_b|y_a,\tau) = p(y_b-y_a,\tau) = \frac{1}{\sqrt{4\pi D \tau}} e^{-(y_b-y_a)^2/(4D\tau)}
\label{eqn:gaussian}
\end{equation}
This probability density satisfies a diffusion equation
with diffusivity $D$. 

It is a remarkable fact that the probability of the maximum excursion
along the path, which is a probability distribution on the space of {\it paths}, can
be determined in terms of the probability of the endpoint given
in Eq. \ref{eqn:gaussian}. This is a consequence of the {\it reflection principle},
discussed in \citet{KS91,PileOuFace}, and holds for a broad class
of random-walk processes. For the unbounded Brownian motion, the maximum
excursion probability is found to be simply
\begin{equation}
P_{max}(y_{max}|y_b,\tau) =
  \begin{cases}
       2p(y_{max}-y_b,\tau) &\text{for $y_{max} \ge y_b$}\\
       0 &\text{otherwise}
  \end{cases}
\label{eqn:NoWallExcursion}
\end{equation}
where $p$ is the probability defined in Eq. \ref{eqn:gaussian}.
We know that the maximum position visited in the past can be no less than the
position at which the particle finds itself at $t=0$, which is
why the probability vanishes for $y_{max} < y_b$.  For $y_{max} > y_b$, the 
reflection principle implies that the probability that $y_{max}$ is
the maximum position visited is twice the probability that the
particle {\it originated} at $y_{max}$.  This seems too good to be true,
but the mathematics is incontrovertible.  Somewhat surprisingly,
the most probable $y_{max}$ is the terminal position $y_b$.  In other
words, the most probable situation is that the value of $y$ 
where the particle is found at $t=0$ is the greatest it has visited
over the past time $\tau$.  Going backwards in time, most trajectories
wander to smaller values of $y$ and never come back to the starting
position.  This is a consequence of the domain being unbounded below, which
allows plenty of room in the domain $y<y_b$ for the trajectories to get lost in.  Note,
however, that the probability of the most probable excursion decays
like $1/\sqrt(D\tau)$ as $\tau \rightarrow \infty$, so that trajectories
with $y_{max} = y_b$ become rather improbable.  For back trajectories
of length $\tau$, values of $y_{max}$ as large as $y_b+\sqrt{D\tau}$ are
nearly as probable as the most probable one. 

Back trajectories with $y_{max} = y_b$ are saturated when they reach
the target positon $y_b$, since the particle has never visited a place with lower $q_s$ 
than it has at its destination.  Because the probability of such trajectories
decays very slowly with $\tau$, the mean moisture content of an ensemble of
trajectories landing at $y_b$ also decays very slowly.  To make these ideas
more explicit, let's consider the case $q_s = \exp(-y)$.  In this case,
a trajectory with maximum excursion $y_{max}$ will produce a specific humidity
$q = \exp(-y_{max})$ when it lands at $y_b$. Solving for $y_{max}$ yields
$y_{max} = - \ln(q)$.  In addition, $y_b = - \ln(q_s(y_b))$, whence
$y_{max} - y_b = - \ln(q/q_s(y_b))$.  This can be substituted for the
argument of $p$ in Eq. \ref{eqn:NoWallExcursion}. From this, we learn
that the PDF of $-\ln(q/q_s)$ at a given point $y_b$ is in fact the
PDF of $y_{max}-y_b$ for back trajectories landing at that point.  The
PDF of humidity may be described as a "truncated lognormal" in the
sense that $-\ln(q)$ is normally distributed for $q \le q_s$, but 
the probability is identically zero for $q > q_s$.  

\begin{figure}
\epsfig{file =  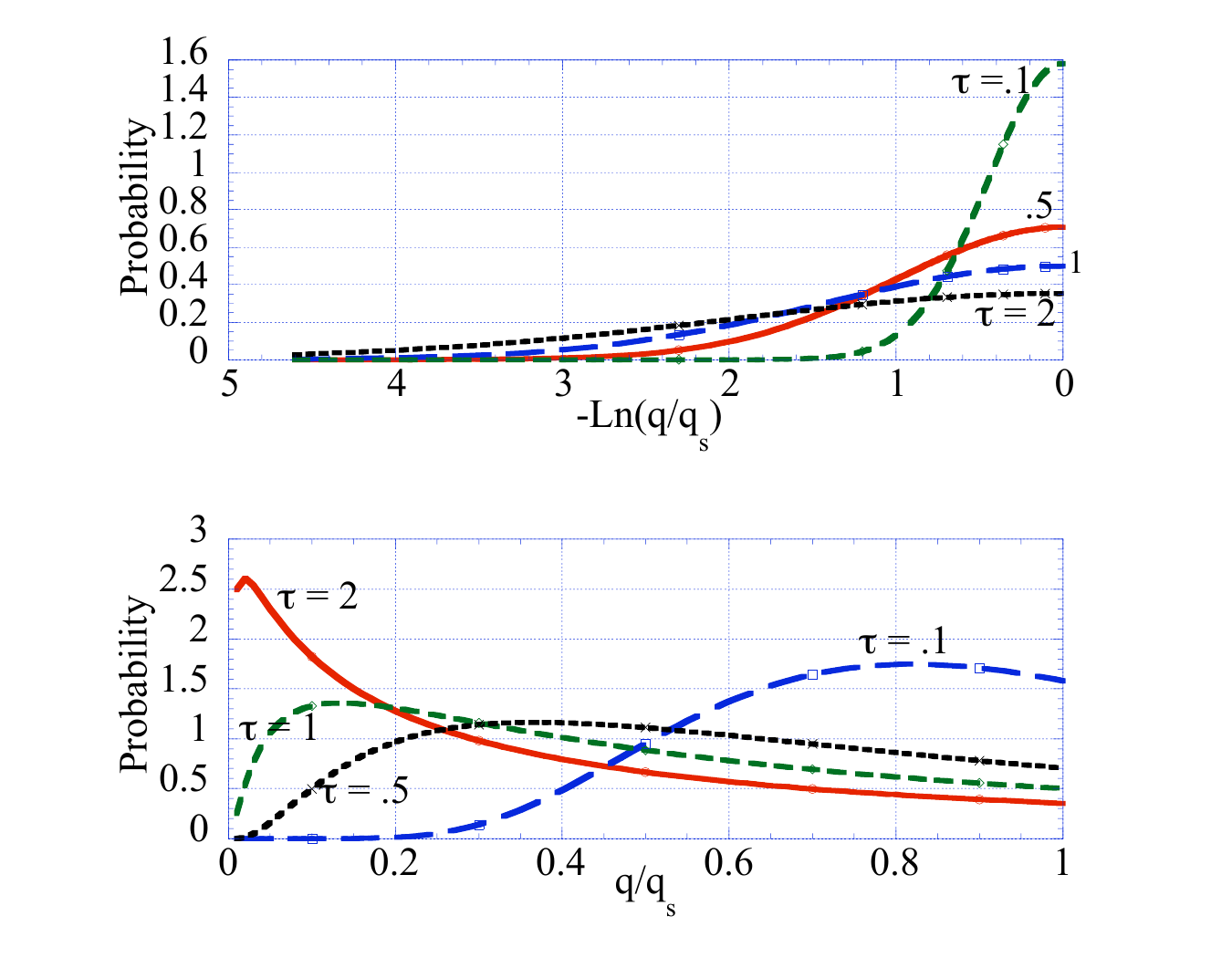, width = 4.5in}
\caption{Probability density functions for $-\ln{q/q_s(y_b)}$ (top panel)
and for $q/q_s$ (bottom panel) for particles executing an unbounded
random walk in a saturation field $q_s(y) = \exp(-y)$. Results are shown
at $t=0$, under the assumption that the entire domain was saturated at the time
 $t = -\tau$ indicated on each curve. Results are shown for $\tau = .1,.5,1,2$ and $2$.
For the exponential saturation profile used in this calculation, the
PDF of $-\ln(q/q_s(y_b))$ is identical to the PDF of the
maximum excursion of back trajectories, i.e. $y_{max}-y_b$.}
\label{fig:NoWallPDFs}
\end{figure}

If we want the PDF of $q$ itself, we must transform the density 
using $-d\ln(q/q_s) = - q^{-1}dq$. Then, letting $Q(q,y_b,\tau)$ be the
PDF of $q$ at point $y_b$, at a time $\tau$ after an everywhere saturated state,
we have
\begin{equation}
Q(q,y_b,\tau) =
  \begin{cases}
       2q^{-1}p(-\ln(q/q_s(y_b),\tau) &\text{for $q \le q_s(y_b)$}\\
       0 &\text{otherwise}
  \end{cases}
\label{eqn:NoWallQ}
\end{equation}
The behavior of $P_{max}$ and $Q$ is shown in Figure \ref{fig:NoWallPDFs}.
Note that while the probability of $y_{max}-y_b$ (or equivalently. of
$-\ln(q/q_s)$) has its peak at zero, corresponding to saturated air,
the peak probability of $q$ shifts towards dry air as time progresses.
This is a consequence of the $q^{-1}$ factor involved in transforming
the probability density from a density in $\ln(q)$ to a density in $q$.  
The population shifts toward unsaturated air as time progresses, though
there is a long moist tail owing to the persistent high probability of
trajectories that do not visit cold places.  The stochastic model
generates dry air in the unbounded domain, whereas the diffusion-condensation
model does not.  The difference in the two models lies in the fact that
the latter represents moisture by a single mean value at any y, and therefore
cannot reflect the fact that amongst the ensemble of particles making up
such an average, some have visited very cold places in the past, and
therefore have become very dry. 

%Figure: PDFs vs time for no wall case

The mean moisture for the ensemble of particles landing at $y_b$ is
obtained by carrying out the integral 
\begin{equation}
\langle q \rangle = \int_0^{q_s} qQ(q,y_b,\tau)dq =  \int_0^{q_s} 2p(-\ln(q/q_s(y_b),\tau) dq
\end{equation}
At large $\tau$, $p$ becomes nearly constant outside of an infinitesimal
interval near $q = 0$, which contributes nothing to the integral in
the limit of $\tau \rightarrow \infty$. Since $p \sim 1/{\sqrt{4\pi D \tau}}$ in
this limit, we conclude that $\langle q \rangle \sim 2 q_s(y_b)/{\sqrt{4\pi D \tau}}$ at large
times.  The humidity in the unbounded random walk model exhibits a slow
algebraic decay, owing to the high probability of back-trajectories wandering off
to regions with large $q_s$ and never returning.  The fact that the moisture
decays at all nonetheless stands in stark contrast to the diffusion-condensation
model in an unbounded domain, for which $q = q_s(y)$ is an exact steady state,
and the air remains saturated for all times.

%Side remark:In general, might be
%best to look at PDF(logq) instead of rh or q, especially in view of radiation effects.

The unbounded case is instructive, but it is very unlike the real
atmosphere, because in the real atmosphere the temperature (and hence saturation specific humidity)
is strictly bounded above by the maximum values prevailing at the tropical
surface.  Real air parcels do not have the liberty of wandering off into
arbitrarily warm regions.  The random walk model can be made more
realistic by imposing a reflecting barrier at $y = 0$, which causes
the saturation specific humidity to be bounded above.  The maximum
excursion statistics for a bounded random walk of this type
can also be obtained by application of the reflecting principle,
but the answer comes in the form of an infinite sum of shifted
Gaussians, which is rather tedious to work with.  The results
we present here are based instead on Monte-Carlo simulation,
with ensembles of 10,000 particles or more.  The PDF does not
sense the presence of the barrier until such time that $\sqrt{D\tau}$
becomes comparable to $y_b$, and particles have had enough time to
frequently reach the barrier.  For longer times, the chief effect
of the barrier is to shift the most probable maximum excursion
to values greater than $y_b$,which moreover increase without bound
as time progresses.  It can be shown that the maximum excursion
scales with $\sqrt{D\tau}$ at large $\tau$. The easiest way to see this
is to note that the trajectories with a reflecting barrier at $y=0$ are
identical to those in an unbounded domain, transformed by reflecting the
negative-$y$ part of the trajectories about $y=0$. The only trajectories
where $y_{max}$ remains small are those that stay near the barrier, and
these become improbable relative to farther-wandering trajectories as time goes on.

\begin{figure}
\epsfig{file =  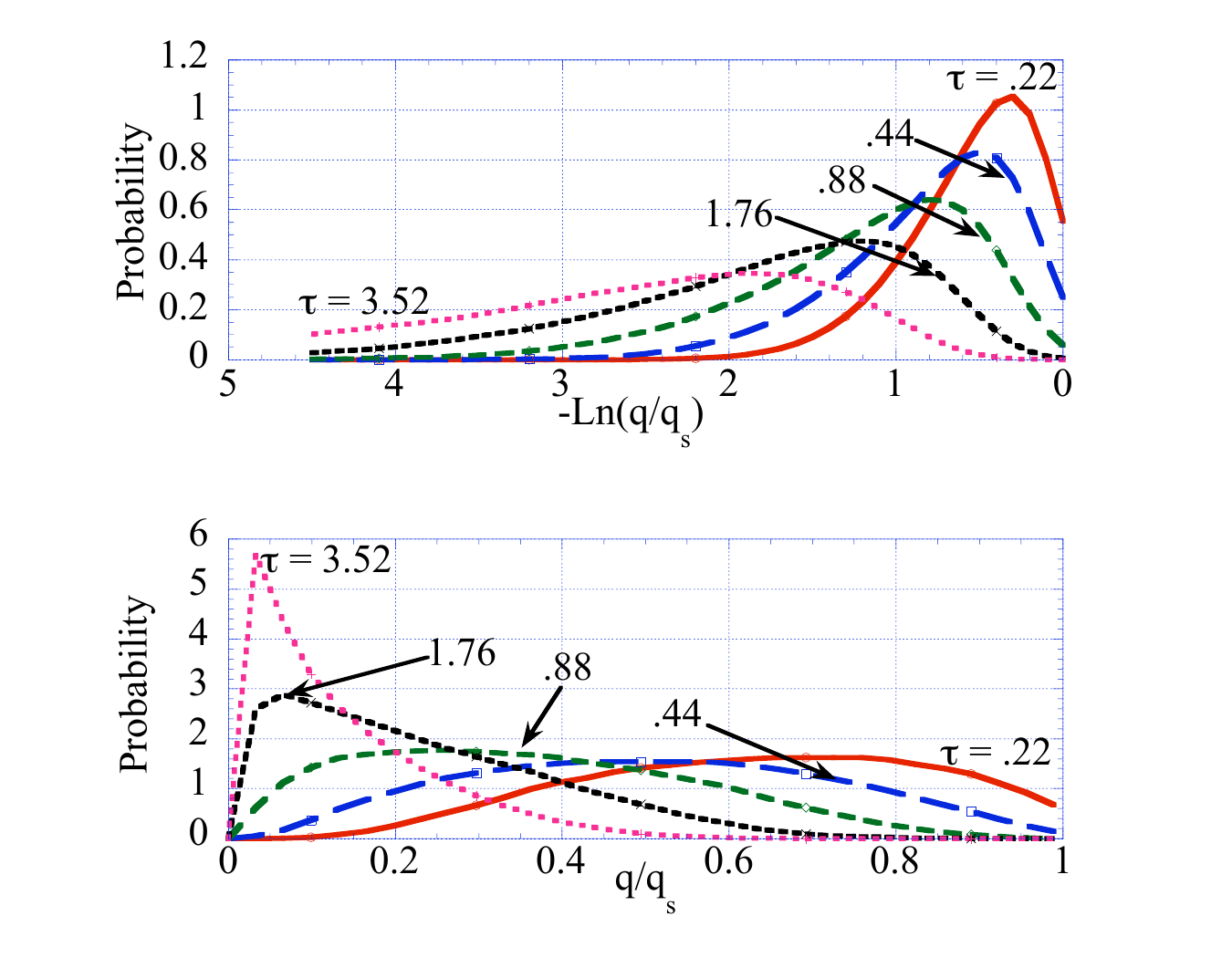, width = 4.5in}
\caption{As for Figure \ref{fig:NoWallPDFs}, but for the case with a reflecting
barrier at $y=0$. In this case, the results depend on $y_b$. These PDF's
were computed at $y_b = .5$. Results are shown for $\tau = .22,.44,.88,1.76$ and $3.52$. 
As before, the top panel can be regarded as the PDF of $y_{max}-y_b$.}
\label{fig:OneWallPDFs}
\end{figure}

The expected behavior can be clearly seen in the upper panel of Figure \ref{fig:OneWallPDFs}, 
where we show the time evolution of the PDF of $y_{max}-y_b$ computed at
$y_b = .5$.  Recall that this PDF is identical to the PDF of $-\ln{q/q_s}$
if $q_s(y) = \exp(-y)$.  Thus, with a barrier, the most probable
value of $-\ln{q}$ shifts towards drier values as $\tau$ increases.  In the
lower panel of the figure, we show the PDF of $q$ itself.  As in the no-barrier
case, the peak shifts toward dry values as $\tau$ increases. However,
the pronounced moist tail disappears after a moderate time has passed,
and nearly-saturated air becomes extremely rare.
%Figure: PDFs vs time for one wall case

\begin{figure}
\epsfig{file = 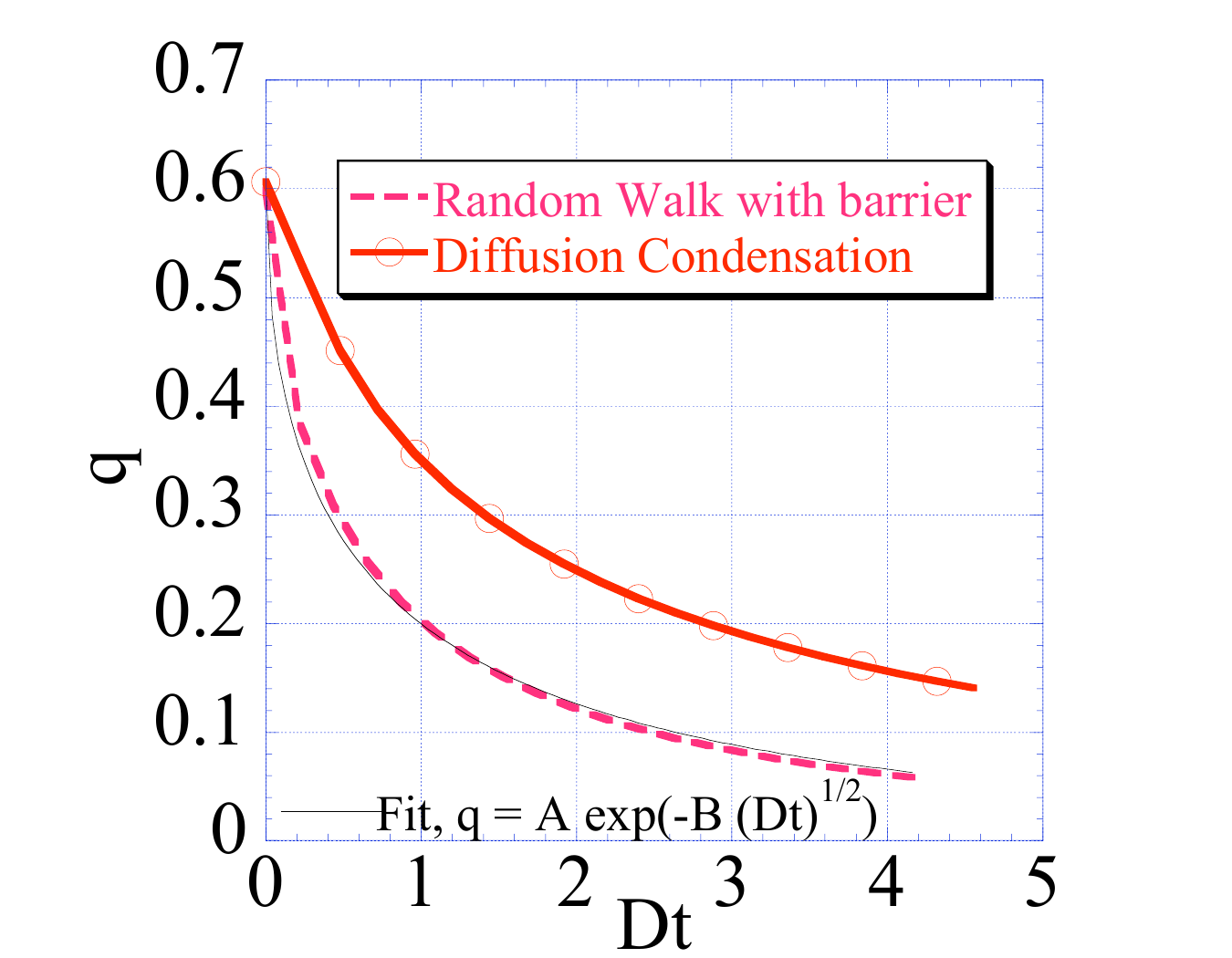, width = 4.5in}
\caption{Decay of ensemble mean specific humidity at y = .5 for the
bounded random walk with a barrier at y=0.  The thin black curve gives
the fit to a stretched exponential decay of the form $\exp(-B \sqrt{Dt})$
The upper thick,solid curve gives the moisture decay for the diffusion-condensation
model computed in the same domain, and with the same diffusivity. The 
saturation specific humidity profile is $q_s(y) = \exp(-y)$}
\label{fig:OneWallDecay}
\end{figure}

The difference with the no-barrier case shows up even more strongly
in the decay of ensemble mean humidity, shown in Figure \ref{fig:OneWallDecay}.
Instead of the slow algebraic decay found produced by the unbounded random walk,
the bounded case yields a rapid decay of the form $\exp(-B \sqrt{Dt})$, in
accordance with the fact that the most probable maximum excursion increases
with time.  In Figure \ref{fig:OneWallDecay} we also compare the stochastic
result with the moisture decay yielded by solving the diffusion-condensation
equation subject to a no-flux boundary condition at y=0. The diffusivity was
chosen so as to correctly reproduce the rate of spread of a cloud of particles
in the random walk case.  We see that the stochastic process dries air
much faster than the corresponding diffusion-condensation process. The roots
of this difference lie in the nonlinearity of condensation. 
%Figure: PDFs vs time for one wall case

\begin{figure}
\epsfig{file = 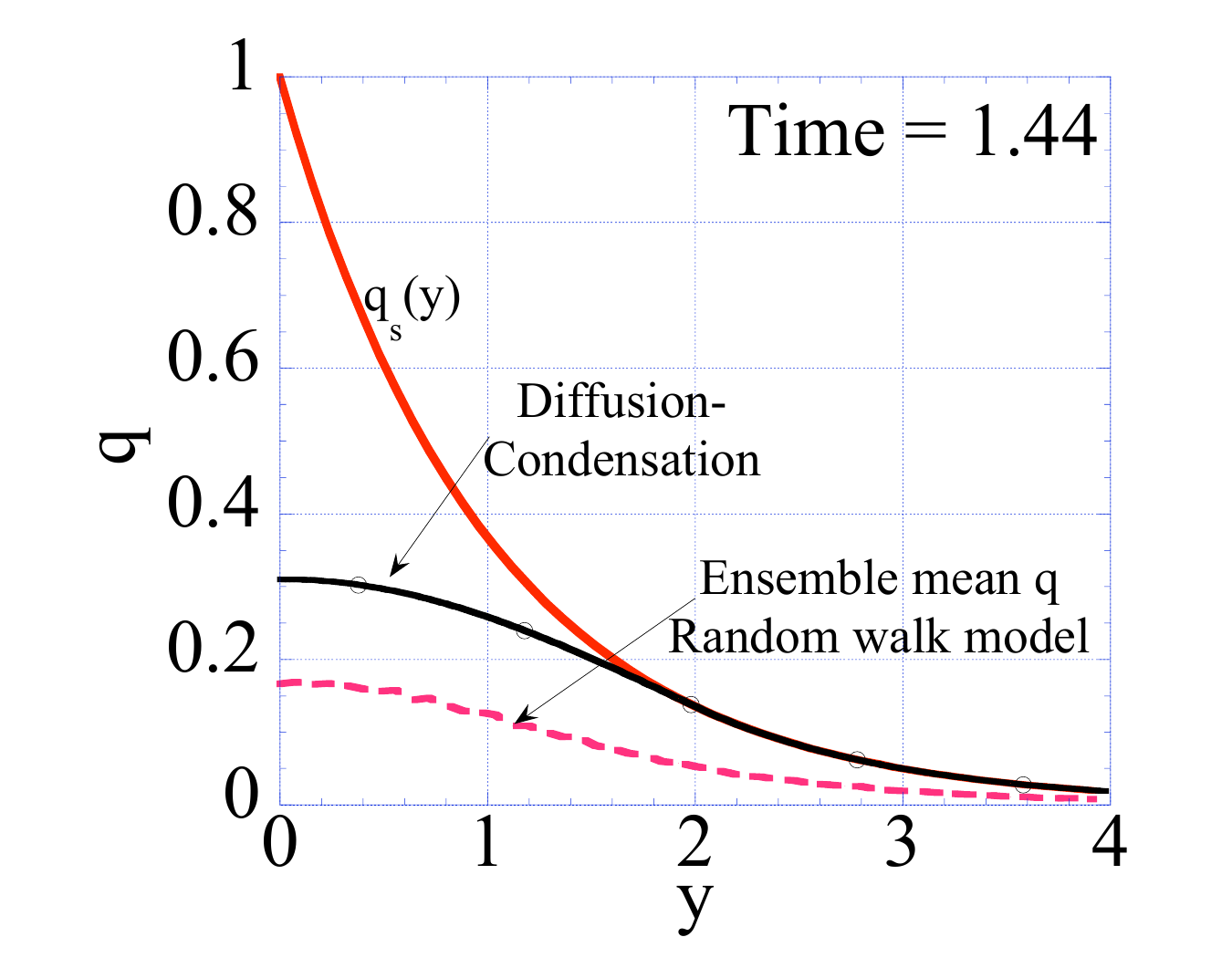, width = 4in}
\caption{Profile of ensemble-mean $q$ at time $D\tau = 1.44$,
for the random walk model with a barrier at $y=0$. The corresponding
profile for the diffusion-condensation model is shown for comparison.}
\label{fig:ACvsRanwalk}
\end{figure}

To shed further light on the comparison between diffusion-condensation
and the random walk model, we show the profile of the ensemble
mean moisture at a fixed time in Figure \ref{fig:ACvsRanwalk},together
with the corresponding moisture profile from the diffusion-condensation
calculation.  At every $y$, the stochastic process yields drier
air than diffusion-condensation.  This happens because some of the
particles in each ensemble are significantly moister than the mean, and
can therefore lose water by condensation upon being slightly displaced.
We can also see an important distinction already familiar from the unbounded case:
the stochastic process generates subsaturated air {\it in situ} even in
a region with $d^2q_s/dy^2 >0$, whereas such regions remain saturated
in the diffusion-condensation model, until such time as dry air invades
from smaller values of y.  The stochastic model thus lacks the sharp
moving front separating condensing from noncondensing regions.  
 
%Figure: Comparison of random wal model with advection-diffusion

%Two wall case doesn't add much new, so just ignore it (driest air
%becomes slightly less dry, that's all).  

To those familiar with the derivation of the diffusion equation in
terms of random walk processes, it may come as some surprise 
that the random walk model of humidity has different behavior
from the diffusion-condensation model. 
If $q$ is a passive tracer, the ensemble average value obtained at each point
by running a series of Brownian back-trajectories to some initial
time does satisfy a diffusion equation. 
Indeed, this technique has been used to obtain Monte-Carlo solutions to
the diffusion equation evaluated along aircraft tracks, without the need
to solve the diffusion equation globally \citep{Legras2003}.
Condensation destroys the means by which the diffusion equation for
the ensemble mean field is derived, precisely because the operation of coarse-
graining (taking the average of concentrations from many trajectories)
does not commute with the operation of condensation.  A saturated and
dry air parcel, when averaged together, will not condense until
subjected to substantial cooling.  The same two air parcels, tracked
separately, will yield condensation immediately on the slightest cooling,
because one is saturated.

The difficulty which emerges from the condensation process is in fact
endemic to all systems where the tracer evolves according to some
nonlinear process, including most chemical reactions.  The difficulty
of describing the evolution of the coarse-grained concentration field
by means of a partial differential equation calls into question
the very notion of an 'eddy diffusivity' when nonlinear processes and
unresolved small scale fluctuations are involved.

We suggest that the random walk model with a barrier provides a 
minimal conceptual model for the generation of dry air in the atmosphere.
It generates a rate and a profile; it generates unsaturated air 
in the interior of regions with exponentially decaying saturation
specific humidity, without waiting for dry air to invade from the boundary;
it predicts the probability distribution of subsaturation, which is
a necessary input to radiation and other nonlinear processes; it has
much in common with the way the real atmosphere generates dry air.  
Thus, it is a much better conceptual model than diffusion-condensation.
As we saw in Section \ref{sec:DiffCon}, when the diffusion-condensation
model is supplied with moisture from a boundary layer that (like
the observed one) is nearly saturated, it saturates the entire
atmosphere except for a thin strip near the ground.  In the next
section, we will see that the random walk model is free from this
shortcoming.

\subsection{Forced equilibria in the stochastic problem.}
\label{sec:StochasticEquilibrium}

\begin{figure}
\epsfig{file=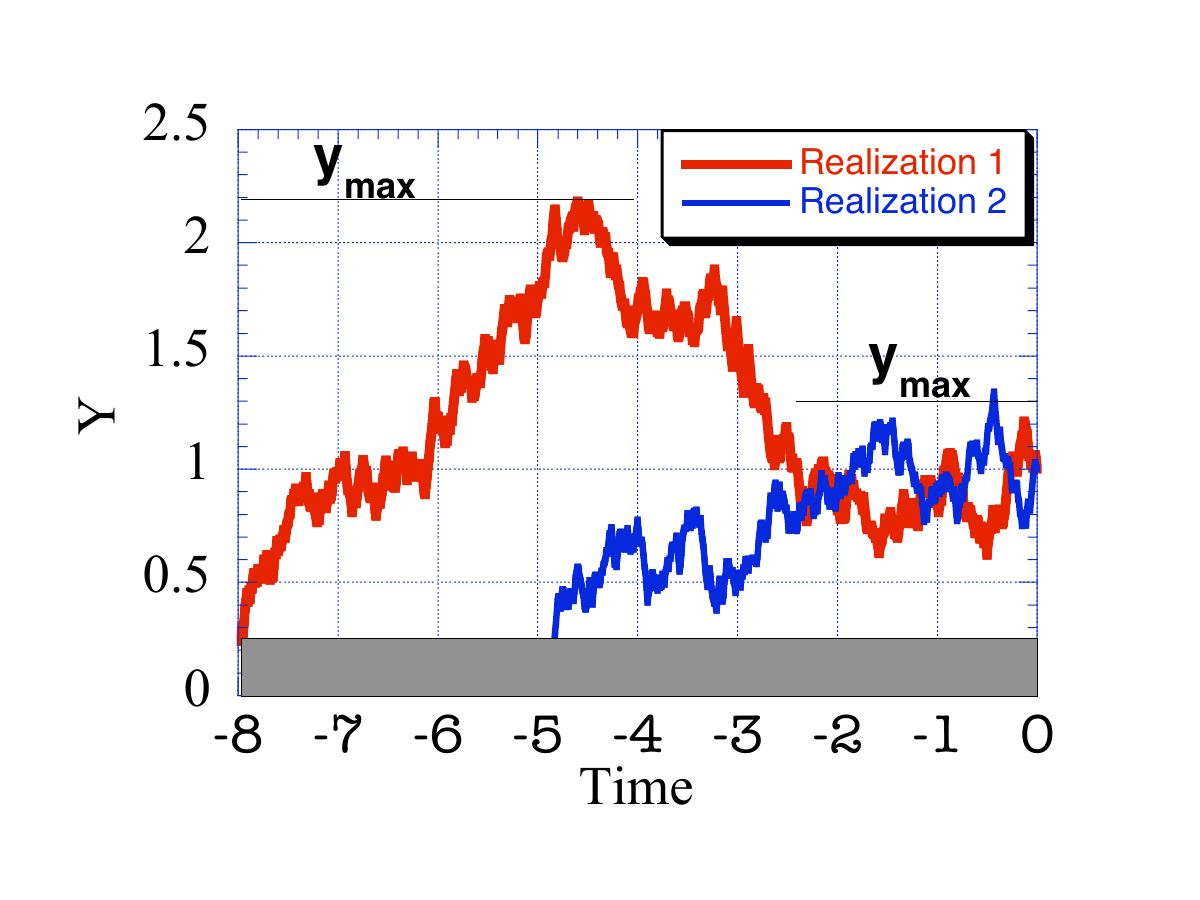,width=4in}
\caption{Two realizations of a random walk between the time of most recent
encounter with the moisture source at the boundary, and the time of arrival
at the target position.}
\label{fig:RanwalkBdd}
\end{figure}

We now examine the statistical equilibria obtained by balancing the
drying processes of the preceding section against some simple
models of the moisture source.  As a first example, we introduce
a moisture source into the bounded random walk model by supposing that
parcels are reset to saturation when they encounter the boundary
at $y=0$.  At large times, the humidity at the target point is no
longer determined by the maximum $y$ (minimum $q_s$) encountered over
an arbitrarily long time $\tau$. The statistic of interest
is now the maximum $y$ attained since the {\it most recent} encounter
with the boundary.  This statistic is illustrated in Figure \ref{fig:RanwalkBdd}.
Back trajectories that take a long time to hit the boundary have a higher
probability of significantly overshooting the target position, and
therefore yield drier values. 

The equilibrium specific humidity PDF is given by Eq \ref{eqn:MoistureConvolution},
where the moistening probability $P_{sat}(\tau)$ is the probability
that a trajectory starting at $y$ first encounters the boundary after time $\tau$.
This encounter time probability is also a classical statistic studied
in the theory of Brownian motion, where it is sometimes referred to as
a "stopping time problem." It can be derived from a maximum {\it negative} excursion
statistic analogous to Eq. \ref{eqn:NoWallExcursion}, giving the probability that the 
{\it smallest} values visited was $y_{min}$, which necessarily is less than
$y_b$.  There is a characteristic time $\tau_D = y_b^2/D$ in the problem.  For
$t \ll \tau_D$, few particles have had time to encounter the boundary, because
the width of spread of the cloud starting at $y_b$ is smaller than the
distance to the boundary.  For $\tau \gg \tau_D$, most particles have already
encountered the boundary, and the probability of a new first encounter
decays like $\tau^{-3/2}$. When plugged into the convolution Eq. \ref{eqn:MoistureConvolution},
these results imply that the equilibrium moisture PDF looks something like
$Q_{min}(q|q(y_b),\tau_D)$, that is, like a freely decaying drying process that
has only been allowed to run for a time $\tau_D$.  However, because of
the fat $\tau^{-3/2}$ tail of the encounter time PDF, there is also a considerable
additional population of dry air, arising from trajectories that have taken
much longer then $\tau_D$ to encounter the boundary.

\begin{figure}
\epsfig{file=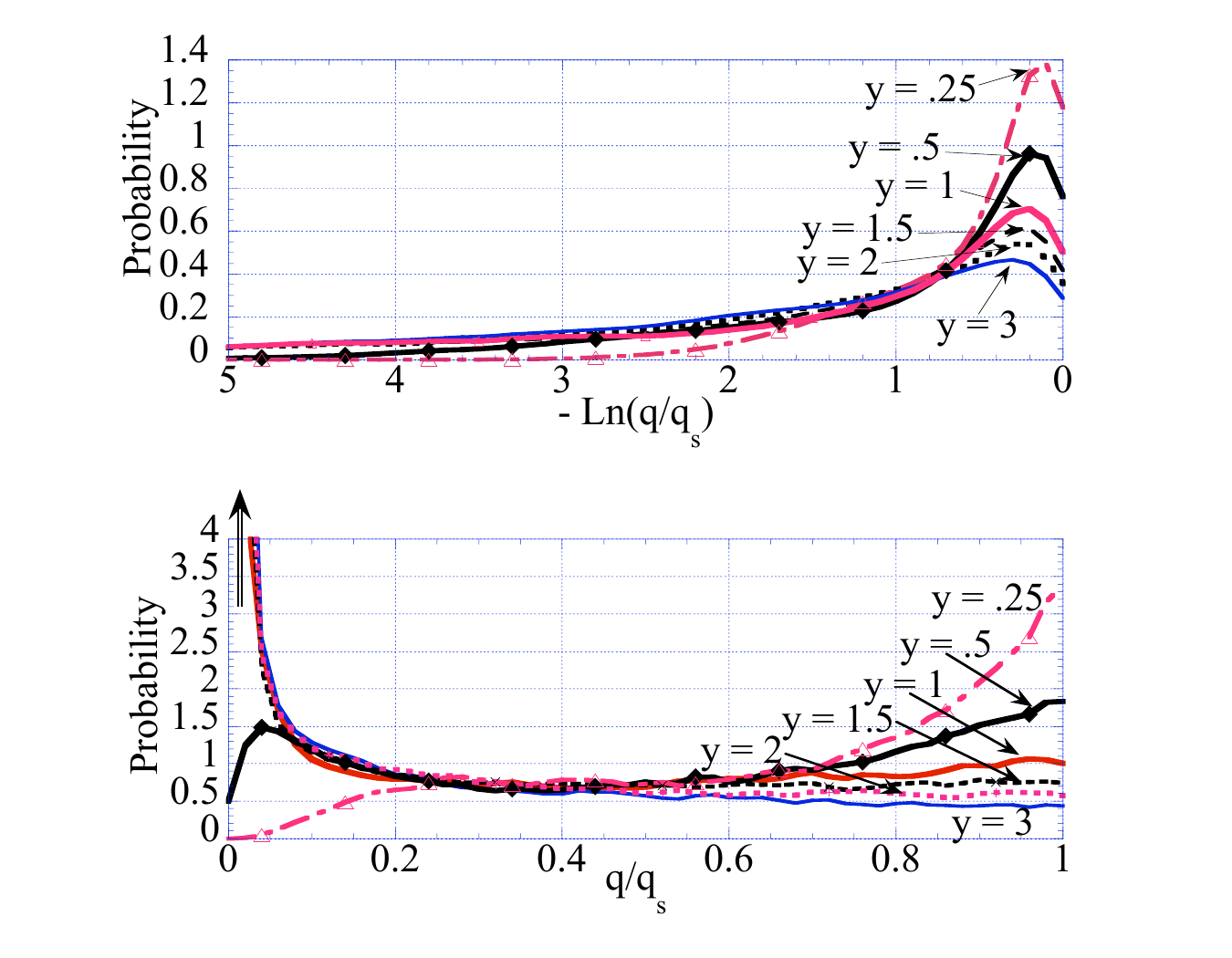,width=4.5in}
\caption{Equilibrium probablitity distributions for $-\ln(q/q_s)$ (top panel)
and $q/q_s$ (bottom panel), at the various positions indicated on the curves.
The PDF's are shown at $y = .25,.5,1,1.5, 2,$ and $3$.
Particles execute a random walk in the domain $[0,\infty]$, and the moisture
tagged to each particle is reset to saturation when they encounter the boundary
at $y=0$.  Calculations were carried out with $q_s(y) = \exp(-y)$. Plotting of the
full height of the dry spike in the lower panel has been suppressed, 
to make the remainder of the behavior more visible.}
\label{fig:eqPDFs}
\end{figure}

Results of an equilibrium Monte-Carlo simulation are shown in Figure \ref{fig:eqPDFs}.
In this simulation, 20,000 particles started at various $y_b$ are random-walked
while keeping track of their moisture, until almost all have reached the boundary.
The moisture PDF's were computed assuming $q_s = \exp(-y)$, as before.  Recall
that in this case, the PDF of $-\ln(q/q_s)$ is also the PDF of the maximum
excursion relative to $y_b$. 
As seen in the top panel, as we increase the distance from the source,
the proportion of particles whose overshoot is near zero reduces, and the
most probable overshoot moves slightly towards larger values. The
most prominent feature at larger distances, though, is that the
distribution develops a fat tail, indicating a fairly high probability
of large overshoots (large values of $-\ln{q/q_s}$).  If $q/q_s$ were
lognormal, the tails in the distribution shown would be Gaussian.
Replotting the data with a logarithmic ordinate (not shown) reveals
that the tails in the equilibrium distribution are exponential rather
than Gaussian. 

In the PDF's of $q$ itself (lower panel),  the fat tails in the
overshoot probability create a pronounced dry spike near
$q=0$.  The overall evolution of the $q$ PDF as $y$ is increased can be
described as a shift of probability from nearly saturated values to 
very dry values.  For positions moderately close to the source, the PDF of
$q$ is distinctly bimodal, with a moist peak and a dry peak; the moist
peak disappears at larger distances.  At all distances, there is a broad
shoulder of intermediate humidities, over which the probability is
nearly uniform.  

%Remark: Note that adding an upper boundary doesn't add much interest
%to the free decay problem, but it is more important for
%the equilibrium problem. An upper wall affects the encounter
%time statistics, by making it harder for parcels to wander
%off to large Y and not get moistened.  Still, the effect
%is not too pronounced, since such parcels tend to have been
%dried a lot since their encounter with the boundary anyway.
%Suppose the boundary is at the tropopause: then any particle
%that "sees" this boundary has already been reset to the (very dry)
%tropopause value, so effect of boundary on the moisture isn't too consequential.
%However: Note that adding an upper boundary does affect the PDF of the
%relative humidity at the upper levels.  The air there is dry, but it can
%still be near saturation (whereas it isn't in the case without an upper lid)
%Mention this, but don't present detailed results. That should be relegated
%to a paper where we seek more realism in stochastic models

\begin{figure}
\epsfig{file=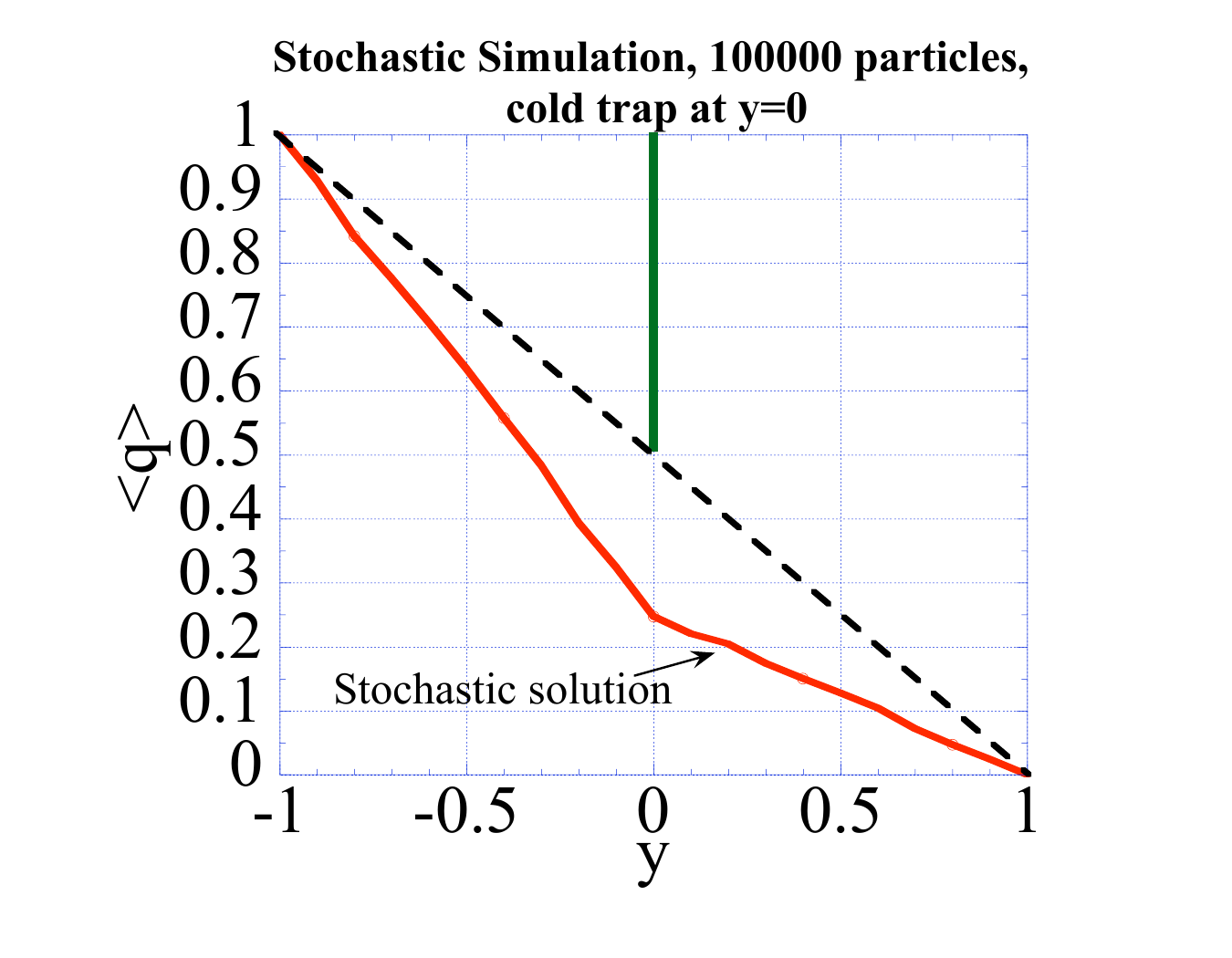,width=4in}
\caption{The equilibrium profile of mean humidity for a cold trap configuration
subjected to boundary conditions $q=1$ at the left hand boundary and $q=0$ at the
right.  The cold trap is at the center of the domain, and resets the moisture to
$q= \frac{1}{2}$.  The dashed line gives the result for the diffusion-condensation
equation, whereas the solid line gives the result of a Monte-Carlo simulation of
the random walk model.}
\label{fig:ColdTrapStochMean}
\end{figure}

\begin{figure}
\epsfig{file=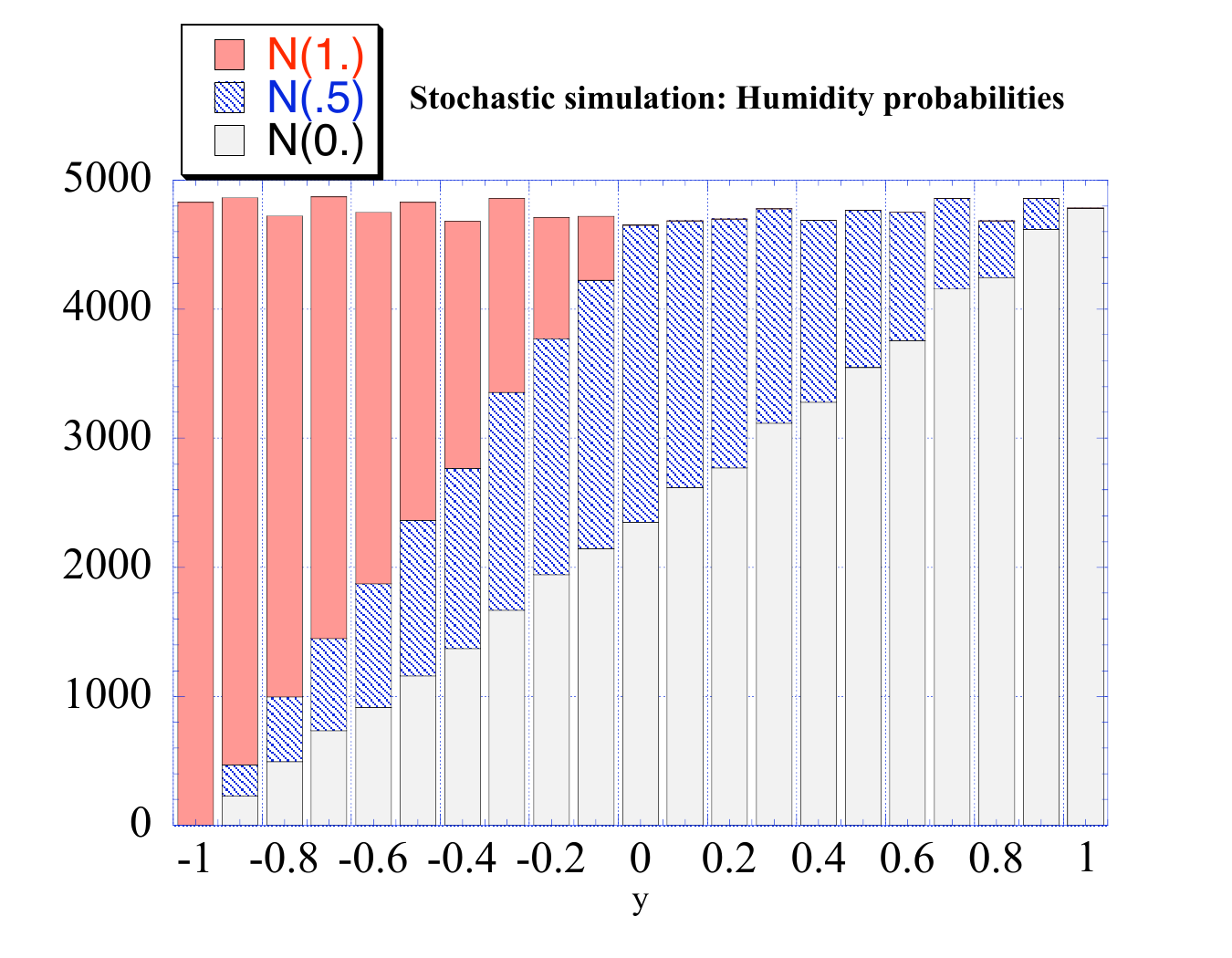,width=4in}
\caption{Spatial dependance of the number of particles of each type in the random
walk simulation of the cold trap configuration shown in Figure \ref{fig:ColdTrapStochMean}.}
\label{fig:ColdTrapStochProb}
\end{figure}

The cold trap configuration  provides a very clear-cut example of the
contrast between mean field theories like diffusion --which model everything
in terms of the field of an ensemble average quantity -- and models
like the random walk model which retain information about fluctuations.
The specific configuration we consider is subjected to boundary conditions
$q = 1$ at $y = -1$, and $q = 0$ at $y = 1$. The saturation specific humidity
is unity everywhere, except for a narrow region near $y=0$ where
$q_s = \frac{1}{2}$.  For the diffusion-condensation model, the equilibrium
profile of $q$ is the straight-line pure conduction profile joining the
boundary values, indicated by the dashed line in Figure \ref{fig:ColdTrapStochMean}.
The cold trap causes no condensation, and has no effect on the moisture profile
in the diffusion-condensation model.  We also solved this problem using the random
walk model, by random-walking 5000 particles, resetting their moisture values
to 1 or 0 upon encounters with the left or right boundary, and resetting the moisture
to $\frac{1}{2}$ upon encounter with the cold trap.  The ensemble mean moisture
profile for this model is given by the solid line in Figure \ref{fig:ColdTrapStochMean},
and has a kink indicating dehydration by the cold trap. It is as if the cold trap
exerts an "action at a distance" caused by the fluctuations about the mean moisture,
allowing it to have an effect on moisture that cannot be determined on the basis
of the knowledge of the mean alone. 
The origins of this behavior are no great mystery.  In the stochastic model,
there are only three possible values for $q$: 1, $\frac{1}{2}$, and 0.  A particle
with $q=1$ will always lose moisture upon crossing the cold trap, so there are no
such particles to the right of the cold trap.  Moreover, particles with $q=1$ to
the left of the cold trap but not too far from it have a high probability of crossing
over and back again, reducing the population of such particles to the left
of the cold trap.  This behavior is illustrated in Figure \ref{fig:ColdTrapStochProb},
which shows how the population of particles depends on $y$.  A population consisting
of half $q=1$ particles and half $q=0$ particles would not condense on the basis
of the average $q$ for the ensemble, but does nonetheless lose moisture since each
individual $q=1$ particle will condense when crossing the cold trap.

What we have just presented is but one instance of a large class of possible
stochastic water vapor models kept in equilibrium by a moisture supply. 
Many refinements or variations are possible. For example, one can include
a second reflecting barrier, at $y_b > 0$; this would
represent some of the effects of the confinement of trajectories within
the troposphere.  Another variant is the random-walk/steady-subsidence
model, which approximates the tropical situation. 
In this case, we think of $y$ as latitude; particles execute
a random walk in $y$ through a horizontally homogeneous $q_s$ field,
but $q_s$ is allowed to vary with pressure $p$ and the trajectories undergo
a steady subsidence in $p$ as they random-walk in $y$.  Parcels are reset
to saturation when they encounter the boundary at $y=0$, which is thought 
of as the region of the tropics which is maintained near saturation by
deep convection.  This model is essentially equivalent to the 
subsidence model described by Eq. \ref{eqn:SubsidencePDF}, with
$P_{sat}(\tau|y_o)$ taken to be the probability distribution of time required for
particles starting at $y=y_o$ to encounter the boundary.  Since the mean
waiting time gets longer as distance from the boundary increases, the air
at any given $p$ becomes drier with distance from the boundary, because
it has subsided more in the time it takes to get there. Still a further refinement
of this model would be to replace the assumption of resaturation at $y=0$ with
an assumption that the parcels are resaturated when they encounter a more general,
spatially complex, set $\Sigma$ in $(x,y)$ space. One would correspondingly
replace the 1D random walk with either a 2D random walk in $(x,y)$, or a random walk
in $y$ and steady sheared advection in $x$.   This model now begins to
approach the model in \cite{P98}, where trajectories were modeled using
observed winds on isentropic surfaces, but with a steady subsidence
across isentropic surfaces, and were resaturated upon encounters with
the actual tropical convective region. 

\section{How will water vapor change in reaction to a changing climate?}
\label{sec:RealTrajec}
It would be fruitful to tinker with the stochastic water vapor model in
search of improved, non-Brownian statistical descriptions of the
trajectories that might better reproduce the salient properties of real
trajectories. The key characteristic of mixing to compute is the maximum
excursion probability distribution. Time-correlated random walks can
still be treated using the reflection principle so long as they are
Markov processes. However, there are few general methods for computing
such path-statistics for non-Markov random walk processes, such as Levy
flights, and it is likely that one would need to resort to Monte-Carlo
simulations to get the needed PDF, if such processes turn out to be
needed to capture the salient characteristics of atmospheric
trajectories. We believe this is the correct path to take in the quest
for water vapor parameterizations suitable for incorporation in
idealized climate models.  In this section, however, we leap ahead to
the direct use of trajectories computed from observed or simulated wind
fields, without any further attempt at a reduced statistical
description.   Our goal in this section is to study how temperature
affects the relative humidity of the free troposphere, taking into
account the Lagrangian nature of the problem.  In particular, we shall
attempt to provide some precise justification for the expectation that
free tropospheric humidity will increase as the climate becomes warmer.

The procedure we employ here is identical to that used in the equilibrium
stochastic model, except that the back trajectories are computed using
simulated or observed/analyzed three dimensional global wind fields.
In particular, the effect of internal mixing and moisture sources are
neglected; trajectories are tracked back to the boundary layer, and assumed
to be saturated at that point.  The moisture at the terminal point of the
trajectory is then the minimum saturation specific humidity experienced 
since the encounter with the boundary layer.  This approach was
employed in \citet{PR98}, who found excellent agreement with the
observed tropical dry zone humidity.  The same advection-condensation
model has since been applied to much larger data sets in 
\citet{Roca2005} and \citet{BrogniezThesis}.  
We will concentrate our efforts on the midlatitude moisture distribution, since
this region has received less attention than the subtropical problem.  
The region of study is the midlatitude Atlantic region bounded by
100E to 0E longitude and 30N to 60N latitude. 

\begin{figure}
\epsfig{file = 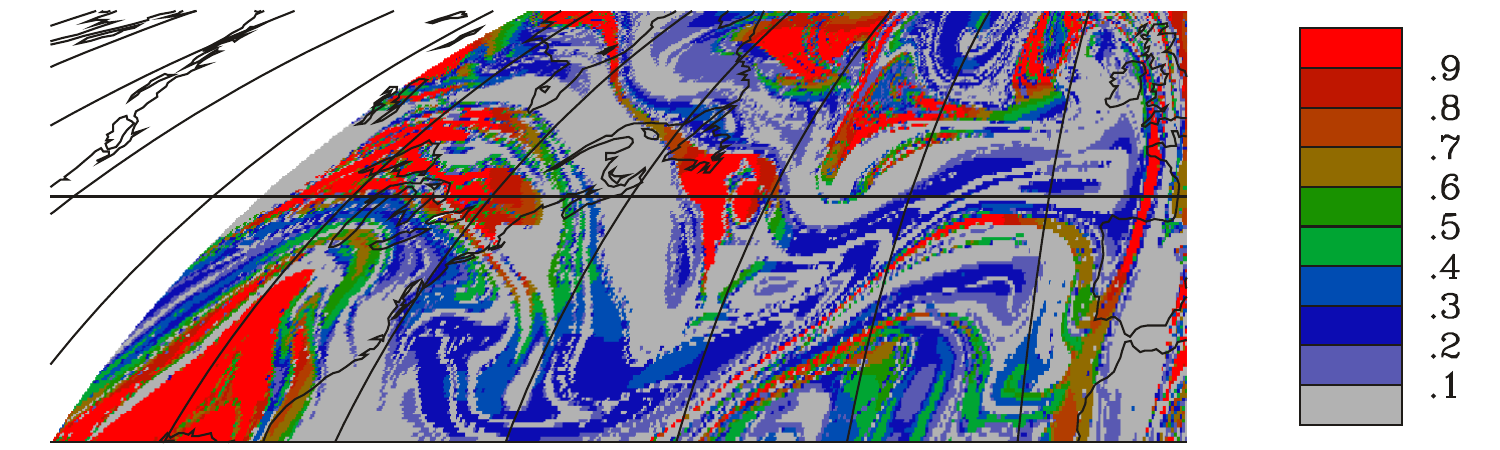,width = 4in}
\caption{500mb Relative Humidity at 12Z Dec 3, 1994, reconstructed
using the back trajectory method driven by NCEP winds and temperature.}
\label{fig:rhmapNCEPTrajec}
\end{figure}

First, we computed the back-trajectory reconstructed relative
humidity fields using NCEP winds and temperatures for December
1994.  The reconstruction was performed four times daily at
a spatial resolution of .25 degrees in latitude and longitude.
To provide a general idea of what the fields look like, a
map of relative humidity at 12Z on 3 December is shown in
Figure \ref{fig:rhmapNCEPTrajec}. It shows the filamentary
structure familiar from earlier work, with small scale
alternations of moist and dry air. 
%**GREYSCALE:Substitute greyscale version of figure
%Figure: Map of rh reconstructed from ncep back trajectories

\begin{figure}
\epsfig{file = 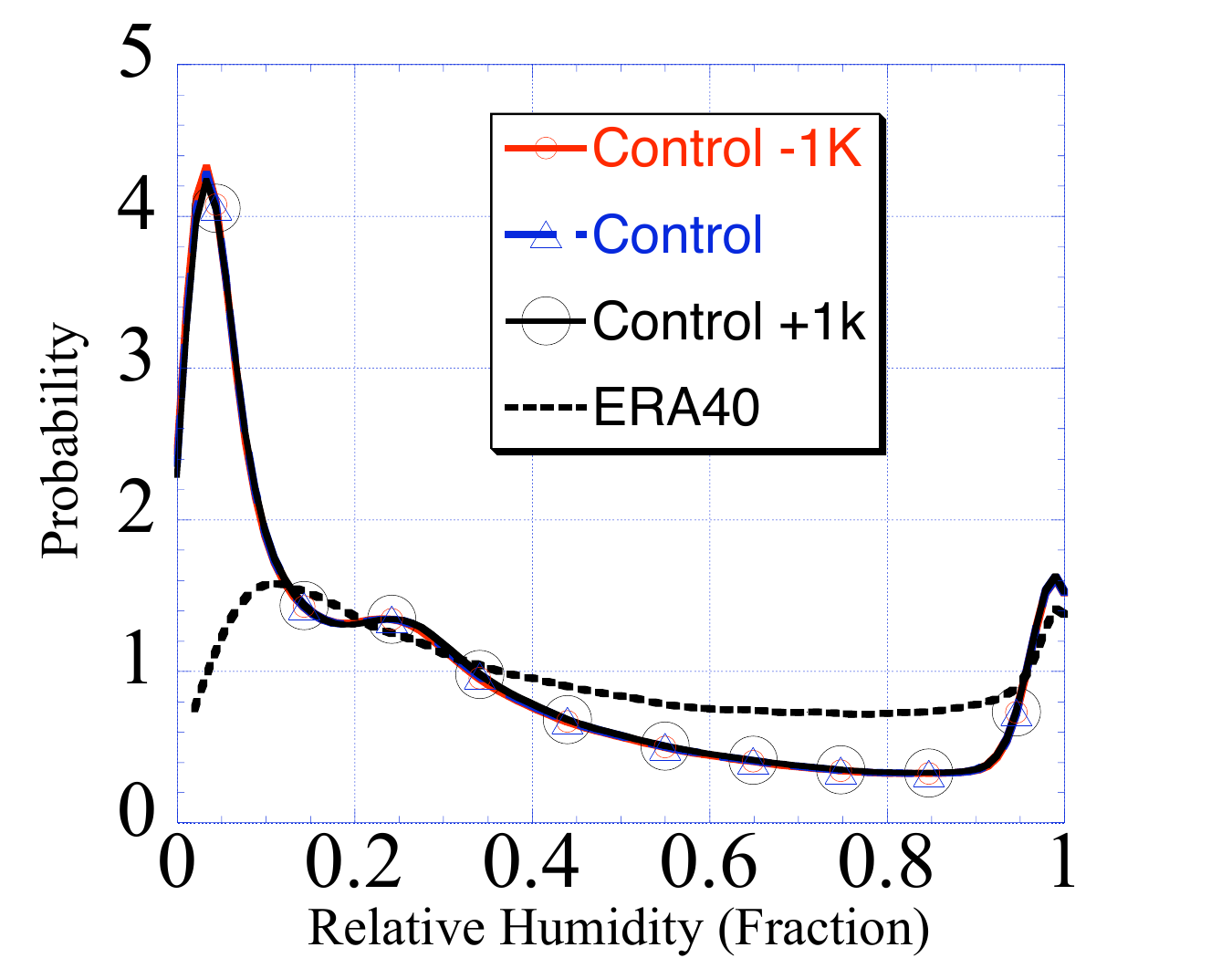,width = 4in}
\caption{Probability distribution of relative humidity for Dec. 1994 over the
region shown in Figure \ref{fig:rhmapNCEPTrajec}, computed
4 times daily using NCEP winds and temperatures.  Results for experiments with temperature
uniformly increased or decreased by 1K are also shown, but the curves are barely
visible because they lie almost exactly on the control case.  For comparison, the
relative humidity PDF over the same time and region for the ERA40 analysis is also shown.}
\label{fig:NCEPTrajecRHPDF}
\end{figure}

The PDF computed over the region of study for the entire month
of data is shown in Figure \ref{fig:NCEPTrajecRHPDF}.  It has a dry spike and
a moist spike, with a broad region of nearly uniform but weakly decreasing probability
in between. In this sense the computed PDF  looks
qualitatively like the PDF for the equilibrium random walk model forced by
a saturated boundary layer, at the point $y=.5$ (Figure \ref{fig:RanwalkBdd}).
For the sake of comparison with an observationally based moisture estimate, we
also show the relative humidity PDF calculated from 500mb ERA40 analyses covering
the same time period and region.
\footnote{We used ERA40 analyses in preference to NCEP analyses because we found
that the ERA40 analyses seem to give better agreement with patterns seen in
satellite retrievals. We used the analyses in preference to satellite data
itself because the latter represent averages over a fairly deep atmospheric
layer, whereas the analyses are available at an individual model level}
As compared to the trajectory reconstruction, the ERA40 PDF has a much less
pronounced dry spike, but a greater population of air with intermediate saturation.
A pattern like this could be obtained by allowing some mixing between moist
and dry filaments in the trajectory model, possibly representing a physical
effect left out of the trajectory model.  It could also result from moistening
of dry air by evaporation of precipitation falling through it. On the other hand,
the lack of a dry spike in the ERA40 result may be an unphysical artifact of
excessive numerical diffusion in the model used to do the analysis/assimilation
cycle. 
Is reality more like ERA40 or more like the trajectory-based reconstruction? We
do not know of any humidity dataset that can unambiguously answer this question
at present. 
%Figure: NCEP traject with \pm 1K, and also ERA40 PDF for comparison

%What about Soden's observation that PDF's are log-normal?  Note that
%the random-walk PDF isn't lognormal in same sense as Soden. Also,
%it is only the cloud-cleared Soden PDF that is log-normal.
%The moist peak seen in ERA40 may be eliminated by the cloud clearing.

Now we pose the question of what happens if the trajectories are kept
fixed, but the temperature at all points in the atmosphere is increased
by a uniform amount $\Delta T$.  In a real climate change, such as
caused by the doubling of $CO_2$, one can expect the statistics
of trajectories to change somewhat, and it is well known also
that the warming is not uniformly distributed in latitude and altitude.
In posing the simplified form of the problem, we are supposing for
the moment that such effects are of only secondary importance
in determining the changes in relative humidity.  The problem
stated this way forms a kind of null hypothesis about the behavior
of water vapor, which can serve as a launching point for more
sophisticated extensions.  The effect of uniform warming on
the relative humidity PDF can be determined by straightforward
reasoning.  Consider a point with temperature and pressure
$(p,T)$ The humidity here is determined by the point $(p_m,T_m)$
where the most recent minimum saturation specific humidity following a resaturation event
occurred (i.e. the position of the "time of last saturation").  
Using the fact that the mixing ratio is conserved in the absence of condensation,
we find that the relative humidity is $(e_s(T_m)/p_m)/(e_s(T)/p) = (p/p_m)(e_s(T_m)/e_s(T))$.
If we increase
temperatures uniformly and keep trajectories fixed, $p_m$ stays the same and the
new relative humidity is $(p/p_m)(e_s(T_m+\Delta T)/e_s(T + \Delta T))$.
Now, if $\Delta T$ is small compared to $T$ and $T_m$, as is usually
the case, the Clausius-Clapeyron relation implies  that the new
relative humidity is approximately
\begin{equation}
rh(\Delta T) \approx (\frac{p}{p_m})
   \frac{e_s(T_m) + e_s(T_m)\frac{L}{R_vT_m^2}\Delta T}{e_s(T) + e_s(T)\frac{L}{R_vT^2}\Delta T} 
    \approx rh(0)(1 + \frac{L}{R_vT} (\frac{T^2}{T_m^2} - 1)\frac{\Delta T}{T})
\end{equation}
This expression assumes the perfect gas law, but otherwise proceeds directly
from Clausius Clapeyron without the need to assume the approximate exponential
form given in Equation \ref{eqn:ClausiusClap}.  This result implies
that the relative humidity increases with warming, but the coefficients
are such that the changes are quite small for moderate values of $\Delta T$.
For example, with $T = 260K,T_m=240K$ and $\Delta T = 1K$, the increase in
relative humidity at each point is only $.014 rh(0)$.  Hence, it is expected
that moderate uniform warming or cooling should leave the PDF of relative
humidity essentially unchanged. 
To confirm this reasoning, we recomputed the NCEP back trajectory PDF's
with temperature uniformly increased by 1K, and also with temperature
uniformly decreased by 1K.  These curves are plotted in Figure \ref{fig:NCEPTrajecRHPDF},
and overlay the control back-trajectory PDF almost exactly.

\begin{figure}
\epsfig{file = 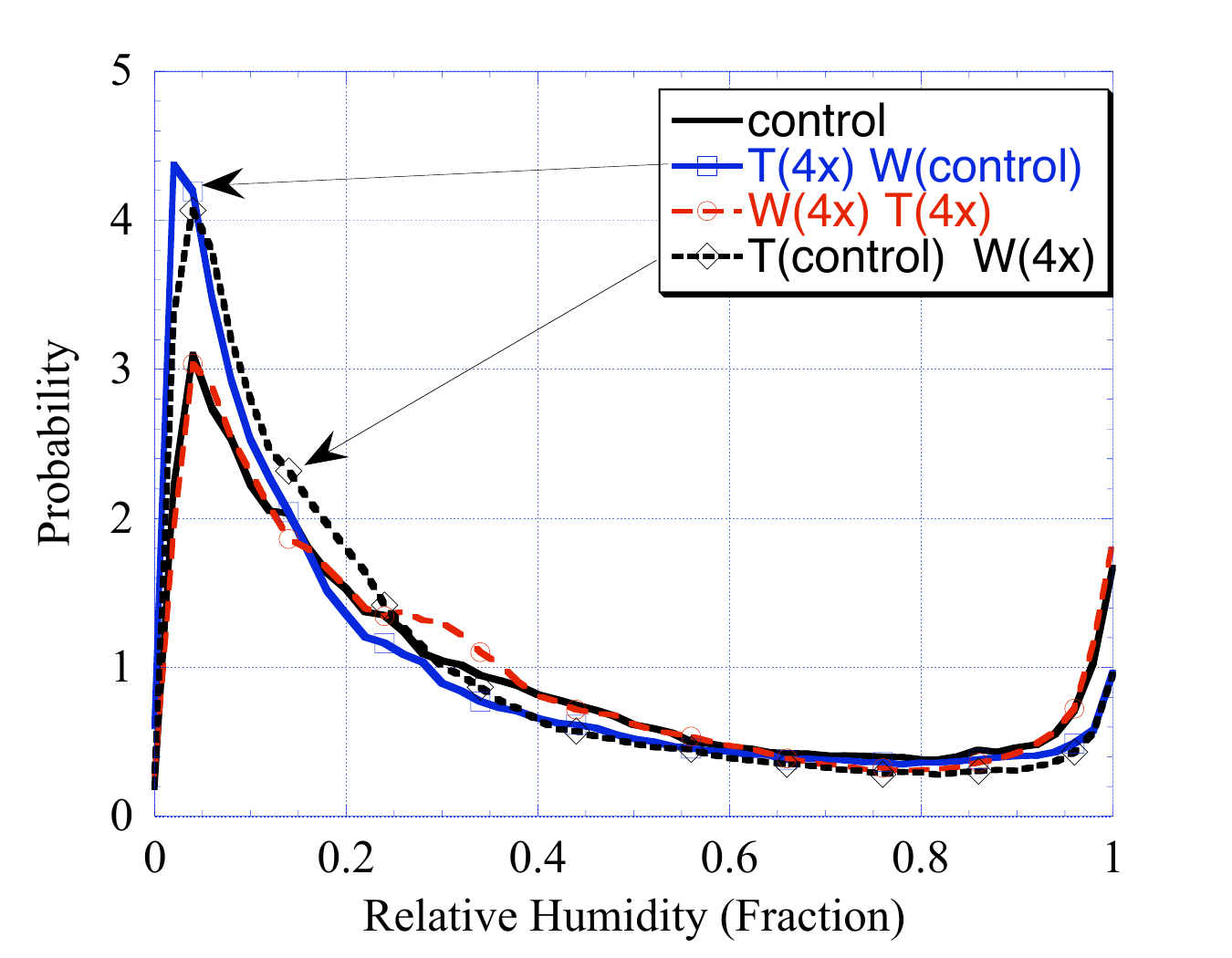,width = 4in}
\caption{Relative humidity PDF's reconstructed by the back trajectory 
technique driven by various combinations of winds and temperature fields from 
a control and a 4x$CO_2$ GCM simulation. See text for definition of the cases.}
\label{fig:FOAMTrajecRHPDF}
\end{figure}

We now apply the same technique to diagnosis of humidity changes in
a GCM simulation of climate change.  This enables us to probe the
effects of changes in the temperature structure, and changes in the
statistics of the trajectories.  We consider two equilibrium simulations
with the FOAM GCM coupled to a mixed layer ocean model.
Two simulations were carried out, both with realistic geography: the first with
300ppmv $CO_2$ (the "control" case), and the second with 1200ppmv $CO_2$ (the "4x$CO_2$" case).  
Figure \ref{fig:FOAMTrajecRHPDF} shows the relative humidity PDFs constructed
using the back-trajectory method applied to the December GCM wind and temperature fields,
for both the control case and the 4x$CO_2$ case. Despite the changes in temperature
pattern and wind fields, the PDF's have almost the same shape for both
simulations. The warm simulation has a slightly greater population of air
with relative humidity in the vicinity of 30\%, at the expense of a reduction
in more saturated air.  We performed two additional back-trajectory reconstructions,
in an attempt to isolate the importance of changes in temperature structure
and trajectory statistics.  The calculation labeled "T(4x) W(control)" was
carried out with the temperature field from the 4x$CO_2$ run but wind fields
taken from the control run.  The calculation labeled "W(4x) T(control)" was
carried out with the wind field from the 4x$CO_2$ run but temperature fields
taken from the control run. Both of these calculations generate more dry air
and less saturated air than the previous two reconstructions.
Changes in the temperature structure in a warmer climate, taken in isolation,
enhance the production of dry air. However, the trajectories in the warmer
climate adjust so as to avoid the colder regions, leaving the PDF invariant
when both effects are applied in conjunction.  
Why this cancellation of effects should take place is at present wholly mysterious.
The cancellation is particularly difficult to understand, in light of the
fact that trajectory changes, taken in isolation, also enhance dry air
production. It will be very interesting to see whether other GCM's exhibit
similar behavior.
%Figure: PDF of RH diagnosed with back trajectory calc driven by FOAM winds and T

\begin{figure}
\epsfig{file = 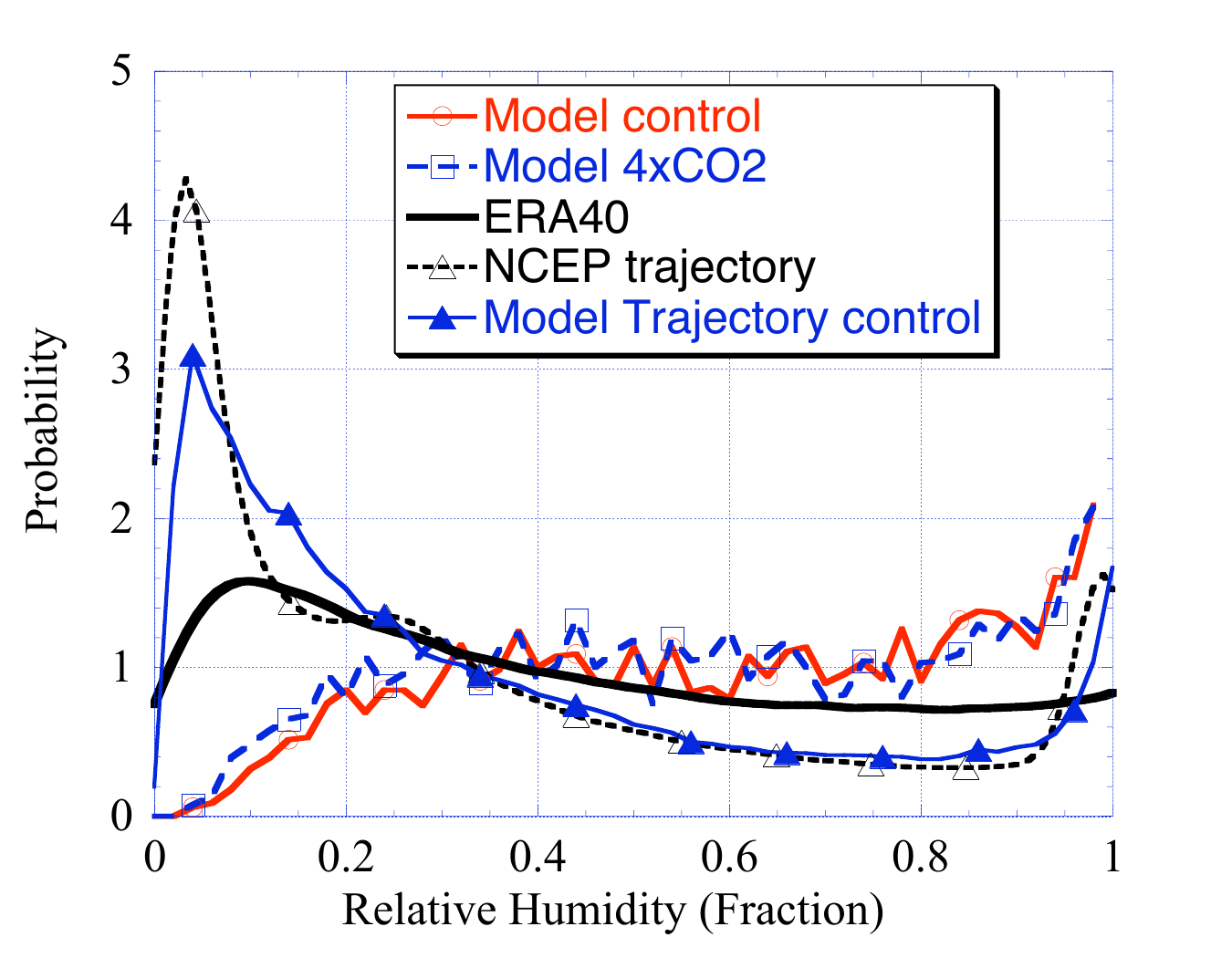,width = 4in}
\caption{Relative humidity PDFs for the internally computed GCM moisture
fields, compared with results from the ERA40 analysis, the NCEP trajectory 
reconstruction, and the trajectory reconstruction driven by GCM simulated
winds and temperature for the control case.}
\label{fig:FOAMRHPDF}
\end{figure}

How does the relative humidity PDF reconstructed from back-trajectories
compare with that based on the humidity computed internally to the model?
These may be expected to be different, since the former reconstructs what
the humidity would be in the absence of mixing and internal moisture sources,
whereas the low resolution GCM allows a considerable degree of mixing. 
As seen in Figure \ref{fig:FOAMRHPDF}, the dry spike seen in both
NCEP and model Lagrangian reconstructions is completely absent from
the internally computed PDF. This is a result of probably excessive
mixing in the GCM.  Significantly, the wind statistics in the GCM do not
seem to be a problem, for the Lagrangian reconstruction based on model
winds is reasonably similar to that based on NCEP winds.  The
PDF from the ERA40 analysis is also shown in the Figure, to underscore
that the GCM is deficient in dry air.  This deficiency is not surprising,
since the leakage of a small proportion of moisture from saturated air is
enough to eliminate a considerable population of dry air.  We do not wish
to imply that this deficiency is characteristic of all GCM's, though it
may indeed be endemic to low resolution GCM's.  It is interesting that, despite
the considerable effect of mixing on the internally computed humidity,
the GCM humidity PDF is nonetheless invariant between the control
and 4x$CO_2$ case. 
%FIgure: Comparison of back trajec. calcs with FOAM model RH, with NCEP back
%trajec, and with ERA40 analysis.

In contrast with the GCM and diagnostic results presented above, the idealized
humidity model of \citet{MD2004} predicts a modest decrease of free tropospheric
relative humidity as the climate warms; this slightly reduces the positive
water vapor feedback, but does not eliminate it.  Since \citet{MD2004}
is a tropical model, whereas the analyses above were applied to a midlatitude
case, a more meaningful comparison must await the application of the
trajectory-based diagnostics to tropical regions.  This is straightforward,
but will be deferred to future work. If the disagreement persists, it might
be due to the neglect of lateral mixing in \citet{MD2004}, or it might be due
to the low vertical resolution of the simulated and analyzed tropical
climate, as compared to the resolution employed in \citet{MD2004}.

\section{Conclusions}

The behavior of water vapor in the climate system is complex and multifold,
and there will probably never be any one simple theoretical framework that
accounts for everything one would like to understand about water vapor.
The problem in its full complexity, as manifest in either atmospheric
observations or comprehensive general circulation models, defies human
comprehension.  Understanding, as opposed to mere simulation, requires
simple models whose behavior can be grasped in its entirety, even if they
are wrong or incomplete in some particulars.  This would
be true even if GCM simulations were perfect, and the necessity is even
more pressing in the face of simulations which are both imperfect and
in certain regards suspect.  We have presented some simple ideas pertinent
to the proportion of subsaturated air in the atmosphere, and the degree
of its subsaturation.  The emphasis on highly subsaturated ("dry") air arises
from its importance in determining the radiative feedback of water vapor.

In particular, based on an analysis of the way large scale atmospheric
trajectories influence subsaturation, we have provided a concise and
defensible statement of why one should expect atmospheric water vapor to
increase as climate gets warmer: {\it The specific humidity at a given point
in the atmosphere is determined by the saturation specific humidity at the
point of minimum temperature encountered along the trajectory extending
backwards in time from this point until it encounters a moisture source
sufficiently strong to saturate it.  If the statistics of the
trajectories do not change too much as the atmosphere warms, this
minimum temperature increases, leading to an increase in the water vapor
content of the target point}. This is a somewhat cartoonish statement
which is modified in its details by closer study, but one that survives
a fair amount of scrutiny. It is intended to replace wholly indefensible
statements which simply invoke the Clausius-Clapeyron relation.
Clausius-Clapeyron is indeed at the root of the behavior of water vapor,
but the proper use of the relation hinges on identifying the
temperature to which the relation should be applied; it's not the
surface temperature, and the effect of the relation on evaporation is of
little relevance to water vapor feedback. 
If one takes the cartoon picture to its idealized extreme, it predicts
that if trajectories are held fixed while the atmosphere warms uniformly
in space and time, the relative humidity probability distribution (PDF) will
remain nearly invariant for small or moderate warming.  
This behavior was confirmed in a midlatitude case, using trajectories
and unperturbed temperatures from NCEP analyses.  

The Lagrangian viewpoint lends itself to the formulation of diagnostics
that can be applied to atmospheric data and GCM simulations, and used
to compare the operation of processes between the two.  In essence, one
uses 3D wind and temperature fields to study the pattern of an
idealized moisture substance which does not diffuse from one air parcel
to another, and which is maintained by an idealized source (typically
in the boundary layer).  Applied to midlatitude NCEP data, the diagnostic yields
a relative humidity PDF whose dominant features are a dry spike
at 5\% relative humidity, a lesser spike at saturation, and a broad
tail of intermediate degrees of saturation in between.  The corresponding
PDF computed from ERA40 data has a much broader and less pronounced dry
peak, centered on 10\% relative humidity, and a greater probability of
air with intermediate humidity.  A key unresolved question at this point
is whether reality looks more like the ERA40 analysis, or the nondiffusive
trajectory reconstruction. The ERA40 analysis, which incorporates
assimilated satellite data, could be indicative of the importance
of mixing processes in the real atmosphere; likely candidates include
vertical turbulent diffusion at small scales \citep{HaynesAnglade}, and
vertical redistribution of moisture owing to evaporation of precipitation
falling through very dry air. On the other hand, the broad dry maximum
in the ERA40 PDF may be symptomatic of excessive numerical diffusion
in the assimilation process.

The Lagrangian moisture reconstruction diagnostic is an instance of the
general idea of studying the behavior of an idealized water-like
substance which isolates some key feature of the problem one would like
to understand, but which is easier to understand than the real thing.
The isentropic study of \citet{YP94} was one of the earlier examples
of this approach. That study examined, in effect, a water-like tracer
that condensed, but had zero latent heat so that condensation would
not cause a trajectory to leave a fixed dry isentropic surface.    
The recent work by \citet{Galewsky2005} also makes use of an idealized
water substance which doesn't release latent heat, but this time in
an Eulerian framework which allows for the effects of mixing amongst air parcels.
In the same spirit, \citet{Frier2006} have used
a synthetic water vapor with adjustable latent heat but no radiative
impact, in order to focus on the influence of moist processes on energy transport.

The Lagrangian moisture diagnostics were also applied to a general circulation
model simulation of warming induced by a quadrupling of $CO_2$. 
The midlatitude relative humidity PDF reconstructed on the basis of trajectories
driven by model winds was found to be invariant between the control
and 4x$CO_2$ cases, even though the model warming is not spatially
uniform and the trajectory statistics are allowed to change.  However,
the invariance of the PDF was found to result from a mysterious cancellation
between the effects of non-uniform warming, and changes in the excursion
statistics of the trajectories, each of which individually causes
a moderate increase in the population of subsaturated air. The
humidity PDF reconstructed on the basis of the model wind and temperature
closely resembles that computed on the basis of observed winds and temperatures.
However, the PDF of relative humidity computed internally to the GCM
is very different from the Lagrangian diagnostic, and is completely
missing the dry peak.  We believe this is probably due to excessive
mixing in the low resolution model. Nonetheless, the PDF of the
GCM internally computed humidity is still invariant under warming,
indicating that the basic invariance inferred from the nondiffusive
Lagrangian calculation survives the addition of a considerable degree
of mixing.

In order to illustrate the manner in which certain trajectory statistics
govern the probability distribution of humidity, we formulated and
analyzed a family of idealized models of water vapor, in which
trajectories are modeled as random walk processes.  The calculation
predicts the shape of the PDF, and the way the
PDF depends on distance from the humidity source.  With some
further refinements, it is even possible that models of this class could
be made suitable for use in idealized climate models, enabling such
models to treat a broader range of questions concerning water vapor
feedback.  Without a moisture source, the stochastic model 
predicts that, in midlatitude conditions, the mean humidity at any given point decays like
$\exp(-\sqrt(Dt)$ for some constant $D$, if the random walk is bounded
above in temperature space.  The humidity PDF
quickly develops a peak at dry values, with an approximately exponential moist tail.  
In the presence of a moisture source at the boundary, the equilibrium PDF
develops a dry spike at sufficiently large distances from the source, with
a "fat tail" extending to more saturated values.  At moderate distances from
the source, the PDF is bimodal, with a secondary peak at saturation. This
shape is similar to that found in midlatitudes using realistic trajectories.

We have compared the predictions of the stochastic model with those of a more
conventional approach based on diffusion of moisture.  Diffusion creates
monolithic regions of saturated air and dries the atmosphere in a manner very
different from the idealized stochastic model, which more closely reproduces
the mechanisms operating in the real atmosphere.  Diffusion is not a suitable
approach to representation of water vapor in simplified climate models, at
least not if one's aim is to treat water vapor feedback.   The comparison between
the stochastic model and the diffusion model also highlights the mathematical
novelty of the former.  Viewed as a stochastic problem, the distribution 
of subsaturation depends on probability distributions on the space of
{\it paths}, rather than just on the space of endpoints of paths encountered in conventional
linear diffusion problems.  Lagrangian history probability problems of this
type cannot be adequately modeled with a mean field theory like the partial
differential equation for diffusion. One needs a Monte-Carlo approach,or
some other approach with retains information about fluctuations as well as ensemble
averages.  This calls into question the whole utility of eddy diffusivity as a means of
parameterizing mixing of tracers subject to a nonlinear removal process.

The trajectory approach, as usually formulated, does not treat regions
of deep tropical convection explicitly. These are detected through the
large scale wind field as regions of persistent upward motion, and
are almost always saturated with moisture.  This has the virtue of
not requiring information about convection that is not implicit
in the wind field, but it does amount to an assumption of saturated
detrainment from convective regions.  This leaves out all the mechanisms
of the sort discussed by \citet{TompkinsEmanuel2000}, which could keep
convective regions significantly subsaturated. The trajectory approach
could be improved by making explicit use of information about convection,
and allowing for subsaturated detrainment. The logarithmic dependence
of outgoing longwave radiation on specific humidity means, however,
that the subsaturation would have to be very substantial before it had
much effect on water vapor feedback.

We have sidestepped the issue of cloud radiative effects by speaking throughout
of "clear sky" radiation when translating water vapor into outgoing longwave radiation.
However, the minute a separation between clear-sky and cloudy-sky radiation is invoked, 
one is implicitly assuming that such a distinction is meaningful.  
The concept of "clear sky" outgoing longwave radiation makes sense 
when applied to large coherent regions free of mid or high level clouds,  and one
should in fact understand the term to allow the inclusion of parts of
the scene including boundary layer clouds (which do not affect the outgoing
longwave radiation). In contrast,  tropical convective regions and the
midlatitude storm tracks are typically characterized by a more or less continuous
and spatially complex distribution of cloudy and clear air, and all types in between.
Since condensed water is so opaque to infrared, the escape of infrared
to space is more controlled by the size of the clear-sky holes between clouds than
by fluctuations in their humidity.  This engages the whole subject of fractional
cloud cover, which may well be the most problematic aspect of cloud representation
in climate models.  Learning to characterize and predict the radiative effect
of intermingled, coupled assemblages of cloudy and clear air is one of the greatest
challenges to the understanding of climate.

\section{Acknowledgements}
The research upon which this chapter is based was funded by the National
Science Foundation, under grants ATM-0121028 and ATM-0123999. I am
grateful to the Editors of this book for numerous valuable suggestions.

\end{document}